\documentclass[]{hdsr}
\graphicspath{{fig/}}
\usepackage{fancyhdr}

\usepackage[utf8]{inputenc}
\usepackage[T1]{fontenc}
\usepackage{hyphenat}
\usepackage{xspace}
\usepackage{amsmath}
\usepackage{amsfonts}
\usepackage{hyperref}
\usepackage{url}
\usepackage{booktabs}
\usepackage{multirow}
\usepackage{makecell}
\usepackage{caption}
\usepackage{minibox}
\usepackage{bbm}
\usepackage{graphicx}
\usepackage{balance}
\usepackage{subcaption}
\usepackage{mathtools}
\usepackage{color}
\usepackage{marvosym}
\usepackage{ifthen}
\usepackage{textcomp}
\usepackage{enumitem}
\usepackage{verbatim}
\usepackage{algorithm}
\usepackage{numprint}
\usepackage{balance}
\usepackage{amsthm}
\theoremstyle{plain}

\newcommand{\abbrevStyle}[1]{#1}

\newcommand{\ie}{\abbrevStyle{i.e.}\xspace}
\newcommand{\eg}{\abbrevStyle{e.g.}\xspace}
\newcommand{\cf}{\abbrevStyle{cf.}\xspace}

\newcommand{\vs}{\abbrevStyle{vs.}\xspace}

\newcommand{\Secref}[1]{Section~\ref{#1}}

\newcommand{\Tabref}[1]{Table~\ref{#1}}
\newcommand{\Figref}[1]{Figure~\ref{#1}}

\newcommand{\xhdr}[1]{\vspace{1.7mm}\noindent{{\bf #1.}}}

\newcommand{\ithdr}[1]{\vspace{1.7mm}\noindent{{\it #1.}}}
\newcommand{\xhdrNoPeriod}[1]{\vspace{1.7mm}\noindent{{\bf #1}}}

\newcommand{\denselist}{ \itemsep -2pt\topsep-10pt\partopsep-10pt }

\usepackage{multirow}
\usepackage{enumitem}
\usepackage{dirtytalk}
\usepackage{amsmath}
\usepackage{amsthm}
\usepackage{amssymb}
\usepackage[utf8]{inputenc}
\usepackage{hyperref}

\usepackage{graphicx, color}

\begin{document}

\newgeometry{bottom=1.5in}

\volumeheader{0}{0}{00.000}

\begin{center}

  \title{In-Class Data Analysis Replications: Teaching Students While Testing Science}
  \maketitle

  \thispagestyle{empty}
  
  \vspace*{.1in}

  \begin{tabular}{cc}
    Kristina Gligorić\upstairs{\affilone,*}, Tiziano Piccardi\upstairs{\affilone,*}, Jake Hofman\upstairs{\affiltwo}, Robert West\upstairs{\affilthree}
   \\[0.25ex]
   {\small \upstairs{\affilone} Stanford University } \\
   {\small \upstairs{\affiltwo} Microsoft Research} \\
   {\small \upstairs{\affilthree} EPFL} \\
  \end{tabular}
  
  \emails{
    \upstairs{*}These authors contributed equally. Work done while at EPFL. Correspondence to: gligoric@stanford.edu. \textbf{Contributions.} Kristina Gligorić: Conceptualization, Methodology, Formal analysis, Writing - Original Draft. Tiziano Piccardi: Conceptualization, Methodology, Formal analysis, Writing - Original Draft. Jake Hofman: Conceptualization, Supervision, Writing - Review \& Editing. Robert West: Conceptualization, Supervision, Writing - Review \& Editing. 
    }
  \vspace*{0.2in}

\begin{abstract}
\noindent
Science is facing a reproducibility crisis. Overcoming it would require concerted efforts to replicate prior studies, but the incentives for researchers are currently weak, as replicating prior studies requires considerable time and effort without providing the same level of recognition as {de novo} research. Previous work has proposed incorporating data analysis replications into classrooms as a potential solution. However, despite the potential benefits, it is unclear what the involved stakeholders---students, educators, and scientists---should expect from it. What are the costs and benefits? And how can this solution help benchmark and improve the state of science?

In the present study, we incorporated data analysis replications in the project component of the CS-401 {Applied Data Analysis} course (ADA) taught at EPFL (École Polytechnique Fédérale de Lausanne), enrolling $N=354$ students. First, we report preregistered findings based on surveys administered throughout the course. We find discrepancies between what students expect of data analysis replications and what they experience by doing them along with changes in expectations about reproducibility. Second, we provide information for educators about how much overhead is needed to incorporate replications into the classroom and identify concerns that replications bring as compared to more traditional assignments. Finally, we discuss potential implications of the in-class data analysis replications for scientific communities, such as insights about replication barriers in scientific work that should be avoided going forward.

Overall, we demonstrate that incorporating replication tasks into a large data science class can increase the reproducibility of scientific work as a by-product of data science instruction.
\end{abstract}
\end{center}

\vspace*{0.15in}
\hspace{10pt}
  
\copyrightnotice

\restoregeometry
\newgeometry{bottom=0.5in}

\section{Introduction}
\label{sec:intro}

The low reproducibility rates of scientific publications have raised concerns across a number of fields~\citep{baker_reproducibility_2016,noauthor_open_2012,open_science_collaboration_estimating_2015}. Although scientific publishing plays a key role in advancing science, the publication process has multiple weaknesses that may influence the validity of conclusions. The focus on novel, confirmatory, and statistically significant results leads to substantial bias in the scientific literature~\citep{thornton_publication_2000}, in fields ranging from basic \citep{begley_reproducibility_2015} and biomedical \citep{goodman_what_2016}, to management and organizational sciences \citep{bergh_is_2017}. This inclination may lead to bad research practices \citep{bishop_rein_2019}, such as $p$-hacking \citep{head_extent_2015,loken_measurement_2017}, or developing post hoc hypotheses to fit known results~\citep{kerr_harking_1998}. 

Recently, \citeauthor{patil_visual_2019} (\citeyear{patil_visual_2019}) introduced a framework to consider the key components of a scientific study pipeline that tend to vary across studies and disciplines: the intent of a study (including research question, experimental design, and analysis plan) and what was actually performed when conducting the study (when data is collected, analyses are conducted, estimates are made, and conclusions are asserted). Replication challenges exist throughout the entire pipeline, all the way to data analysis, given previously collected and publicly available data. Data analysis replication, in particular, entails different analysts using their independently written data analysis code to reproduce the original estimates and claims, using the same data and the same analysis plan~\citep{hofman_expanding_2021}. {Such data analysis replication corresponds to a computational reproduction based on the original data, but without the original code}~\citep{national2019reproducibility}. 

Significant variation in the results of data analysis replication has been proven difficult to avoid, even {when the incentives are well-aligned} \citep{silberzahn_many_2018}. Researchers are increasingly encouraged to share code and materials~\citep{nosek_preregistration_2018} for other researchers to perform direct data analysis replication, as a way to improve the credibility of the corresponding research findings. However, replicating the data analysis reported in the publications of others requires considerable time and effort, without providing a particularly rewarding outcome, that is a publication, because of a presumed lack of originality~\citep{janz_bringing_2016} and novelty \citep{open_science_collaboration_estimating_2015}. Researchers are thus usually not incentivized to perform data analysis replications. Ultimately, published replications are rare across fields~\citep{makel_facts_2014,lemons_inadvertent_2016,perry_decade_2022,plucker_replication_2021,king_replication_1995} and the incentives are not yet in place to address this issue.

A recent body of work~\citep{hofman_expanding_2021,quintana_replication_2021} has proposed one step toward a solution: educating undergraduate and graduate students to perform data analysis replications. Universities are well positioned to introduce replications as class assignments in methods training in order to establish a culture of replication~\citep{mendez-carbajo_data_2023,ball_yes_2023}, reproducibility, and critical thinking \citep{janz_bringing_2016,stojmenovska_teaching_2019,smith_role_2021,chopik_how_2018}. In-class replications have previously been proposed for college-level education \citep{meng_reproducibility_2020} and for psychology education~\citep{hawkins_improving_2018,frank_experimentology_2023}, to understand correlates of replicability \citep{frank_teaching_2012,boyce_eleven_2023}. Furthermore, data analysis replication efforts have previously been used for comprehensive meta-analyses \citep{wagge_publishing_2019, wagge_demonstration_2019}, based on multiple studies rather than on a single replication attempt \citep{boyce_eleven_2023,shrout_psychology_2018,perry_replication_2022}.

However, despite the postulated advantages of this solution, it is unclear what the involved stakeholders---students, educators, and scientists---should expect from it. First, in terms of \emph{students}, it is unclear, what type of effort does this require on their end? What do students learn from the process, that is, how do their beliefs differ before \vs\ after engaging in data analysis replication exercises? What outcomes do students expect before the activity, and how do actual outcomes differ from prior expectations?

Second, in terms of \emph{educators}, there are open questions regarding required investments \vs\ potential advantages over traditional exercises. For instance, what is required on the educator's end to run successful data analysis replications? How can data analysis replications be incorporated into existing large university classes? What should educators expect their students to learn and take away from data analysis replications? How much of the educator's time and effort is in-class replication expected to take, and what challenges might the educator face~\citep{stojmenovska_teaching_2019}?

Lastly, in terms of \emph{scientists}, it remains to be determined how this solution can help benchmark and improve the state of science. What are the main sources of error or confusion that students identify? How can these replication barriers in scientific work be avoided going forward?

To provide new insights about the in-class data analysis replication approach, we incorporated data analysis replications in the project component of the \textit{Applied Data Analysis} course (CS-401) taught at EPFL, the Swiss Federal Institute of Technology in Lausanne.

\xhdr{CS-401 class: Background} {This course taught the basic techniques, methodologies, and practical skills required to draw meaningful insights from data. The course had the following prerequisites: an introduction to databases course, a course in probability and statistics, or two separate courses that include programming projects. Also, programming skills were required (in class, we mostly used Python). Most students were first-semester students in computer science or data science master's programs (although registration was open to students from other programs who meet the requirements). At the start of the class, a typical student had strong programming skills and was familiar with fundamental concepts related to algorithms, computer systems (e.g., databases), and the fundamentals of probability and statistics.}

{During the semester, the students learned the methods during lectures and were introduced, in the lab sessions, to the data analysis software tools. In parallel, the students worked on an applied data analysis project. In a regular iteration, for the project component, students proposed and executed meaningful analyses of a real-world data set. These required creativity and the application of the methods and tools encountered in the course. The outcome of this team effort was a project report and a publicly available code repository.} 

{Lastly, at the end of the semester, students took a 3-hour final exam where they completed a data analysis pipeline on a data set they have never worked with before. By the end of the class, a student is typically able to construct a coherent understanding of the techniques and software tools required to design a data science pipeline.}

\xhdr{Our approach in integrating data analysis replications into CS-401} {As part of the project component of the class, instead of the standard unconstrained data analysis project leveraging a real-world data set, students individually performed data analysis replications.\footnote{{The replication-based project contributed to 25\% of the final course grade (the remaining 75\% was split between the homework, quizzes, and the final exam).}} Class setup was otherwise unchanged, except for adjustments necessary to run the replication exercises (\cf \Secref{sec:costs}).}

Based on a set of surveys conducted over the course of the semester, our main goal was to understand students' expectations about the difficulty of the exercise before performing the replication \vs\ their impressions of how hard the task actually was, once completed. Through preregistered analyses of survey responses, we pose the following specific research questions about the impact that data analysis replications tasks have on the students. Our guiding research question is: How large are the discrepancies between students’ expectations and the reality of data analysis replication, {in terms of time investment, perceived difficulty, tasks, and outcomes} (RQ1)? Additionally, we explore the following questions: Do the discrepancies (if any) persist in subsequent replication tasks, after the first one is solved (RQ2)? Can students anticipate in what ways peer-reviewed data science papers might be hard to replicate (RQ3)? Finally, are the effects stronger for the same type of data analysis as performed in the replication exercise, or is there an attitude shift for expectations regarding different data analysis types (RQ4)?

Any discrepancies between expectations and reality ({RQ1--2}) and any changes in expectations about reproducibility in general ({RQ3--4}) serve as evidence of shifts in students' attitude. Identifying such indicators of behavioral changes is essential to understanding students' experiences of performing data analysis replications.

\begin{figure*}[t]
\centering
    \includegraphics[width=\textwidth]{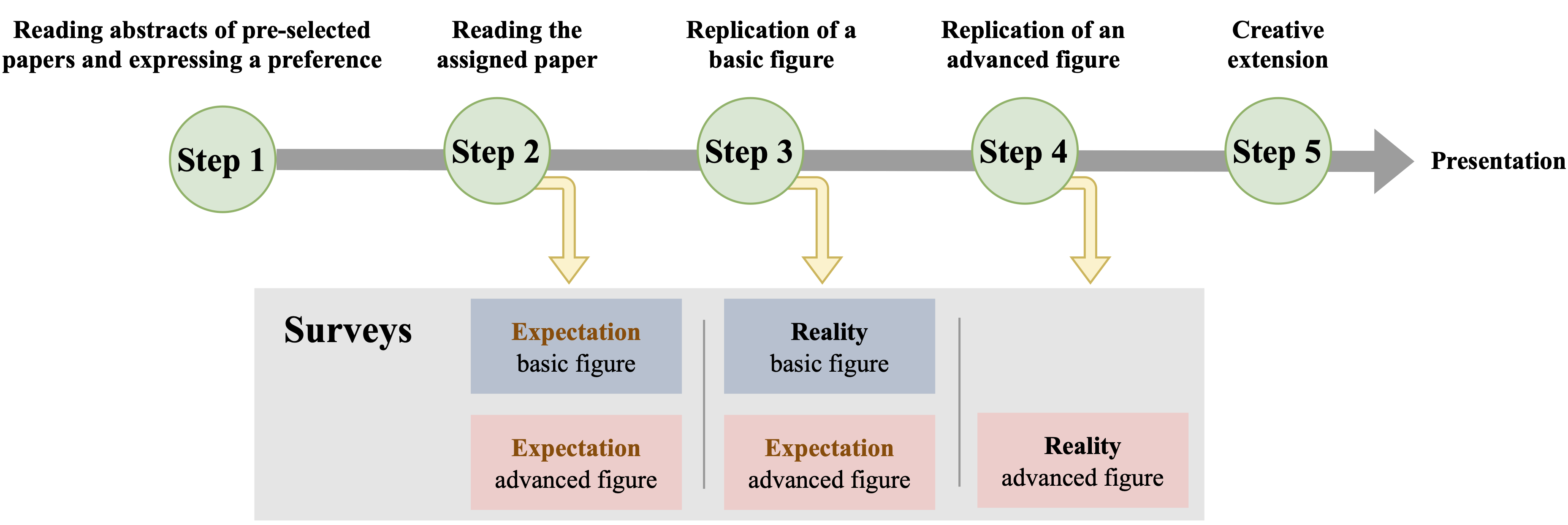}
\caption{\textbf{Study design summary.} The timeline is visualized from the students' perspective. The semester progresses from the left to the right. The surveys were administered upon submission of the respective assignment step.}
\label{fig:study}
\end{figure*}

\xhdr{Overview of study design} The replication activity was performed as part of the graded class project. {Replication exercises were conducted individually.} The study design is outlined in \Figref{fig:study}. The study started with a bidding process where students expressed preferences for papers (Step~1). Afterwards, each student focused on one scientific paper, assigned to them by the class instructors. After reading the assigned paper (Step~2), presurveys recorded the individual students' expectations about the time required, the difficulty of replicating findings from data science papers, and about the perceived reproducibility of papers in the field.

Then, students performed the replications (Steps 3 and 4). Replications were performed and reported individually by each student. We specified two figures or tables to replicate, a basic one (replicated in Step~3) and an advanced one (replicated in Step~4). Students then individually recorded their results and working times in postsurveys, which we compared with students' expectations from before they started as expressed in the presurveys. Lastly, students proposed and conducted creative extension projects, which students built on top of the replicated analyses (Step~5) and presented at the end of the class.

\xhdr{Contributions} {Concretely, we describe in-class data analysis replication and report `lessons learned' as relevant for students, educators, and scientists. Our findings are based on the work and responses of $354$ consenting students who produced data analysis replications of $10$ peer-reviewed publications.\footnote{\url{https://dlab.epfl.ch/teaching/fall2020/cs401/reports/}} Moreover, creative replication extensions performed at the end of the class are contrasted with standard, unconstrained projects, conducted the following year.}

\ithdr{Students} {In total, 98\% of students reported having replicated exactly or qualitatively the basic figure, and 87\% the advanced figure. A small fraction of replications failed, and in cases where there were known issues with papers, students correctly identified them. We found that it takes students on average about 10.5 hours to replicate a main result (\cf\ \Secref{registered}), {and further 8.5 hours to replicate the second result (19 hours in total)}. Discrepancies between expectations and reality, and changes in expectations about reproducibility arose among students.\footnote{Preregistration: \url{https://osf.io/usm4k/}}
On average, students underestimated the time they would take to reproduce, overestimated how long data wrangling would take, and underestimated how long it would take to iteratively analyze and interpret results (\Secref{registered}).}

{The identified attitude shifts signal students' enhanced appreciation for the challenges involved in the scientific process. Exploratory analyses of open-text responses (\cf\ \Secref{exploratory}) then let us identify how the students perceived this activity and understand the specific challenges that the students faced, including resource, expertise, and time constraints. Further exploratory analyses of creative extensions on top of data analysis replication show that replication extensions might be both more methodologically advanced and scientifically meaningful than unconstrained projects conducted the following year.}

\ithdr{Educators} {On the educator side, we provide realistic information about how much overhead is needed in teacher-to-student ratios for overseeing replications, how much effort is required to select papers for replications, and some concerns that replications bring over more traditional assignments. We offer further `lessons learned' that can be useful to other educators, putting particular emphasis on reflections regarding cost--benefit tradeoffs. The insights about the discrepancies between expectations and true outcomes, as well as the associated attitude shifts, will be informative for future efforts aiming to incorporate data analysis replications into existing educational practices. For example, since the replication activity took students longer than expected, instructors should carefully plan the course timeline and clearly communicate the expected workload to students, to avoid stress and frustration (\cf \Secref{sec:costs}).}

\ithdr{Science} {Lastly, we identified potential ways how the scientific communities could benefit from this and similar efforts. Overall, we demonstrated that incorporating replication tasks into a large data science class has the potential to increase the reproducibility of scientific work as a by-product of data science instruction.} 


\section{Methods}
\label{sec:methods}

\subsection{Study Design}

In preparation for the study, we identified 10 data science publications suitable for the course, in terms of the difficulty of data analysis tasks required, a variety of covered topics, and data availability. The publications were split into five tracks, with two publications each: 
\begin{enumerate}
\denselist
\item Natural language processing and machine learning \citep{niculae_linguistic_2015,muchlinski_comparing_2016}
\item Computational social science \citep{choi_predicting_2012,pierson_large-scale_2020}
\item Networks \citep{leskovec_signed_2010,cho_friendship_2011}
\item Social media and Web \citep{penney_chilling_2016,liang_testing_2015}
\item Health \citep{aiello_tesco_2020,cattaneo_housing_2009}
\end{enumerate}

We identified two key figures or tables from each of the publications that are important for the overall message of the publication. Teaching assistants (master's students who took the course the previous year) aimed to replicate (exactly or qualitatively) the selected figures before the class started, {which ensured that the selected figures and tables were qualitatively reproducible}. We developed pre- and postsurveys by conducting a pilot with student assistants.

The data analysis replication activity was composed of six steps. We asked students to fill out repeated short surveys, each part of a project milestone deadline. Each student was assigned one paper to replicate (around 36 students per replicated paper). In each paper we selected a primary and secondary figure or table. The primary figure or table requires basic skills taught in the lectures and exercises before the replication task was performed (limited to counting, hypothesis testing, visualizing, and fitting regressions). The secondary figure or table requires potentially more advanced data analysis such as nonstandard resampling and error estimation techniques, examination of feature importance, and network analysis. Note that henceforth we refer to the basic figure/table and the advanced figure/table as simply \emph{basic} and \emph{advanced figures} (although the result might be presented in a table).

Additionally, the 10 assigned papers were divided into two conditions based on the paper type (referred to as type A and type B). Paper type refers to the type of analysis necessary to perform the replication. For basic figures in type A papers, to reproduce a result, students were required to count items, perform hypothesis testing, and make a visualization and interpretation of the result (papers: \cite{niculae_linguistic_2015}, \cite{liang_testing_2015}, \cite{cho_friendship_2011}, \cite{aiello_tesco_2020}, and \cite{leskovec_signed_2010}). For type B papers, to reproduce a result, students had to fit a regression model and make a visualization and interpretation of the result (papers: \cite{choi_predicting_2012},\cite{pierson_large-scale_2020}, \cite{cattaneo_housing_2009}, \cite{muchlinski_comparing_2016}, and \cite{penney_chilling_2016}). In addition to the main assigned paper (referred to as `Paper 1'), each student was assigned two control papers (referred to as Papers 2 and 3) that the student does not replicate. One control paper was of the same type as the replicated paper, and one of the other type.

The study consisted of five steps, outlined in \Figref{fig:study}:

\xhdr{Step 1: Reading abstracts of preselected papers and expressing a preference} Students were instructed to read abstracts of all the 10 preselected papers to get an idea of what the papers are about. Students then ranked the 10 papers by their preference of working on them for the project. After this, students were assigned a main paper (`Paper 1'). We assigned the same number of students per paper. We calculated the average rank of preference for each paper across the students, and assigned papers to students in a balanced way, to minimize the total average rank since smaller rank implies higher preference. We also assigned to each publication two assistants who were in charge of mentoring students working on the respective data analysis replication.

\xhdr{Step 2: Reading the assigned paper} Students were instructed to read the assigned paper. Students were pointed to the freely accessible PDF and the data set repository. Students wrote a short summary (at most 500 characters). Upon submission of the summary, the students completed the presurvey measuring expectations of the replication of the assigned figure and general attitudes toward reproducibility.

\xhdr{Step 3: Replication of a basic figure} Students individually performed a replication of the assigned basic figure from the assigned paper (`Paper 1'). Students prepared a replication report in the form of a Jupyter Notebook containing {independently written code and text}. Students were instructed to log their hours spent doing the replication, on a piece of paper, in a digital sheet, or using time-tracking software. To elicit truthful time log reports, it was clarified to the students that the number of reported hours would in no way affect the grading of the work. Upon submitting the replication report, the students completed the first postsurvey, which measured outcomes of the replication of the basic figure and expectations for the advanced figure. The main analyses (RQ1) contrast the postsurvey responses after replication of the basic figure with the presurvey responses given before the replication exercise.

\xhdr{Step 4: Replication of an advanced figure} In order to work on other graded assignments in the course, students formed groups of four students (some groups exceptionally comprised three students) working together throughout the semester. In their group, the students then proposed a creative extension of the analysis performed in the paper, placing their data science skills into practice \citep{kolaczyk_statistics_2021}.\footnote{Recall that the group members individually replicated the same paper in the previous steps.} When submitting the short project proposal, the students also completed the second postsurvey, a repeated measurement of the expectations for the advanced figure. Analyses in RQ2 contrast the second postsurvey with the expectation for the advanced figure.

\xhdr{Step 5: Creative extension} Students conducted the proposed creative extension in their group. Additionally, individually and following identical instructions as in step 3, the students replicated the advanced figure from the assigned paper. Students were again instructed to log their hours spent doing the data analysis replication. The students completed the third postsurvey, measuring outcomes of the replication of the advanced figure, and general attitudes toward reproducibility. General outcomes toward reproducibility are studied to address RQ3.

After each step, students were additionally asked about their expectations about the two control papers that they did not replicate (Paper 2 and Paper 3). We explore answers related to these control papers in order to address RQ4. In preparation for the study, we tested this pipeline with five student assistants.

\subsection{Inclusion and Exclusion Criteria} The study took place at EPFL in the fall semester of 2020, between September 2020 and January 2021. In total, 384 students took the class. Out of 384, 30 students (7.81\%) opted out from the study (resulting in $354$ consenting students). Data from all the enrolled students were analyzed in the study, except from those who chose to opt out.
We also excluded students who did not submit all four surveys or whose responses did not pass validation checks.
With these restrictions, we analyzed responses from $N=329$ consenting students.

\subsection{Consent Statement and Information Sheet} Students were provided with the following information about the study and its purpose: \say{As part of ADA 2020, we introduced data analysis replications as a way of making you interact with real data science research. In order to understand the effectiveness of this new learning paradigm, we will analyze your solutions and survey responses, and we aim to publish a research paper about our findings. No personal data will be made public; we will only release aggregate, anonymized information. Every data point is valuable for us, but if you would nonetheless like to retract your data from the analysis, you can indicate this by checking the following box. Checkbox: I would like to be excluded from the analysis of the ADA data analysis replications.} An information sheet about the study was provided to students.\footnote{\url{https://go.epfl.ch/ada2021-replic-info-sheet}}

\section{Results: Data analysis replications}
\label{sec:results}

\subsection{Preregistered Findings: Discrepancies Between Expectations and the Reality of Data Analysis Replication}
\label{registered}

Before analyzing the data collected via surveys, we formed and preregistered a set of primary and secondary hypotheses, each relating to one of the four research questions (RQ1--RQ4).\footnote{Pre-registered data analysis plan and survey materials are publicly available: \url{https://osf.io/usm4k/}} We then executed the analyses following the plan. Our primary confirmatory analysis tests the hypothesis that there are discrepancies between students’ expectations and the reality of data analysis replication (H1 [RQ1]). {An overview of the results is presented in ~\Tabref{tab:h1}.} Replication package including code and data is publicly available~\citep{DVN/A6VMD9_2024}. 

Testing the preregistered hypotheses, we first asked: \emph{Is there a significant difference between the time students take to perform the data analysis replication and the time they expect to take (H1a)?} We found that there is a significant difference ($p=.0309$; full distributions in \Figref{fig:H1a_1} and \ref{fig:H1a_2}). On average, students expected to take 9.01 hr, but actually took 10.53 hr. The median expected time is 5 hr and median time taken is 8 hr. In total, 62\% of students took longer than expected, 7.30\% the same, and 30.70\% less than expected. So overall, students on average underestimate the time it would take to reproduce the basic figure.

Second, we compared how challenging students thought that it would be to reproduce the basic figure from the assigned paper with the reported true level of challenge. Specifically, we asked: \emph{Is there a significant difference between how challenging performing data analysis replication tasks is and how challenging students expect it to be (H1b)?} We found that there is a significant difference (illustrated in \Figref{fig:H1b_1} and \ref{fig:H1b_2})---interestingly, performing data analysis replication tasks was less challenging than expected ($p=3.70 \times 10^{-5}$). The average expected score on the 1--5 scale is 3.39 (median 4), whereas the average score after performing the task is 3.11 (median 3). 

Third, we conceptualized the data analysis replication task as being composed of three core activities: {data wrangling} (understanding the data structure, preprocessing steps, feature engineering), {data analysis} (exploratory analysis, statistical tests, developing and training models, evaluating model performance), and {interpretation} (evaluating results and comparing them with the results in the paper, interpreting findings, and redoing the analysis if necessary). Then, we then asked: \emph{Are there discrepancies between the predicted and the true distribution of time spent on the three core activities: data wrangling, data analysis, and interpretation (H1c)?} We found discrepancies ($p<10^{-307}$)---in relative terms, students overestimated how much time data wrangling would take, and underestimated how much time data analysis and interpreting results would take (\Tabref{tab:2}). This finding shines light on why replication took more time than expected, but was less challenging than expected. Students took more time iteratively redoing the data analyses, interpreting their results, which was perceived as time-consuming, although not technically challenging.

Finally, we asked: \emph{Are there discrepancies between predicted and true outcomes of the replication (H1d)?} First, we found that 98\% of students reported having replicated exactly or qualitatively the basic figure, and 87\% the advanced figure. We did not find significant discrepancies between predicted and true outcomes of the replication ($p=.0747$; illustrated in \Figref{fig:H1d_1} and \ref{fig:H1d_2}). A possible explanation is that the papers were preselected to be (with enough effort) at least partially qualitatively replicable. Students were not exposed to randomly sampled papers from the field. Rather, the selected papers were already found to be qualitatively reproducible in our paper selection process. Further statistics are available in the Appendices, Appendix A: Primary Hypotheses--statistics and data distribution visualization.

\begin{table}[t]
\setlength\extrarowheight{-2pt}
\scriptsize
\centering
\renewcommand{\arraystretch}{1.5} 
\begin{tabular}{l|l}
\textbf{Outcome} & \textbf{Summary statistics}\\
\midrule
H1a: Time spent                               & Students underestimate the
time it would take to reproduce: \\
& 1.52 hour increase ($p=.0309$). \\
& Pre test: $M=9.01hr$, post test: $M=10.53hr$.                                                                                    \\
\hline
H1b: Level of challenge                  & Data analysis replication was less challenging than expected:\\

& 0.28 point decrease ($p=3.70 \times 10^{-5}$).                                                                              \\
& Pre test: $M=3.39$, post test: $M=3.11$, on 1--5 scale.                                                                                    \\
\hline
H1c: Time distribution  & 
Students overestimated time for data wrangling, and underestimated time for\\
& data analysis and interpreting results:\\

& Significant disturbance in the ranking ($p<10^{-307}$).\\
& Pre test ranking in decreasing order: Wrangling, Analysis, Interpretation,\\
&post test ranking in decreasing order: Analysis, Interpretation, Wrangling\\
\hline
H1d: Replication outcomes                            & Difference not significant ($p=.0747$). \\
& Pre test: $M=1.81$, post test: $M=1.75$, on a 1--3 scale.  \\  
\end{tabular}
\caption{{\textbf{Results overview: Modified expectations (H1).}} Summary of the results comparing pre- and postreplication expectations, across the four hypotheses (H1a--H1c).}
\label{tab:h1}
\end{table}

We also considered a set of secondary hypotheses (\textbf{H2--4}). First, we hypothesized (H2 [RQ2]) that discrepancies between predictions and true outcomes persist as students solve replication tasks (complete statistics available in Appendices, Appendix B: Secondary hypotheses). Overall, when reproducing the advanced figure after the basic one, discrepancies between expectations and outcomes persisted (although some in the opposite direction). Most notably, there were discrepancies between the predicted and the true distribution of time spent on the core activities and between predicted and true outcomes of the replication, since there were replication failures that the students did not expect. Second, we found no evidence supporting the hypothesis that the replication task affects the students' expectations on the fraction of peer-reviewed data science papers that are reproducible (H3 [RQ3]; Appendices, Appendix B: Secondary Hypotheses).

Third, we hypothesized that there is a spillover effect as expectations are modified across the board, to papers that students did not replicate (H4 [RQ4]). Overall, we indeed found that there is a spillover effect as expectations regarding time spent and time distribution across the activities are modified for the papers that students did not replicate (summarized in \Tabref{tab:h4}). 

\xhdr{Summary} Overall, we found that data analysis replication tasks take longer, but are less challenging than expected. Compared to the expectations, students spent more time analyzing and modeling the data and interpreting the results, and less time in data wrangling activities. We did not find significant discrepancies between predicted and true outcomes of the replication. The considerable amount of time spent modeling and interpreting the results may explain why replication took more time than expected, while simultaneously being less challenging than expected. We found that students took time iteratively redoing the data analyses, and interpreting their results, which was perceived as time-consuming, although not necessarily technically challenging. The identified discrepancies and attitude shifts signal students' enhanced appreciation for the challenges involved in the scientific process.

\subsection{Exploratory Findings: Understanding the Students' Experience}
\label{exploratory}

Next, we complement the previous findings with an exploratory study identifying the challenges students experienced during the replication activity, to understand the gaps between the expectations and the reality of data analysis replication. In this analysis, we qualitatively investigate the open-text responses to two questions we included in the postsurvey: (1) ``What was challenging?'' (2) ``What may explain the differences?''
Students replied to these questions after replicating the second figure and completing the replication assignment.

To understand what topics the students mentioned, two of the authors of this study qualitatively coded the students' answers using a grounded-theory approach. For both questions, we independently repeated the following process. The researchers autonomously read a random sample of 100 answers and produced a list of topics mentioned in the students' descriptions. These topics were then compared and discussed until an agreement on their representativeness was reached. This process led to merging similar topics and refining the names describing them. 
Then, each researcher assigned the obtained topics to the answers. Multiple topics (or none) could be assigned to an answer. Finally, the label assignments were compared and, in case of discrepancies, discussed until a final agreement was reached. At the end of the process, the answers not assigned to any previously agreed topics were examined to extract new possible labels. If new topics were identified, the process was repeated; otherwise, the process terminated by leaving these answers unlabeled. The outlined topic coding approach was applied to two open-ended questions included in the postsurvey: (1) ``What was challenging?'' (2) ``What may explain the differences?'' We report topics assigned to at least 5\% of the answers. 

\xhdrNoPeriod{What was challenging about the data analysis replication?}
In this question, students were asked to describe in two or three sentences what they found challenging during the replication task. Most students (77\%) described challenges assigned to at least one of the topics. Inspecting the unassigned responses (23\%) did not lead to introducing additional themes. Rather, the unassigned responses were short and vague (\eg, \say{probability issue}) or uninformative (\eg, \say{It did not replicate at all}).

We identified four frequent topics:
\textsc{Poor description}, \textsc{Expertise requirements}, \textsc{Time requirements},  and \textsc{Limited resources}. In the following paragraphs, we report more details about the four themes and the relative commonness in the questions assigned to at least one topic. Since each answer can be assigned to multiple topics, the percentages of assignments do not sum to 100\%.

\vspace{3mm}
\noindent \textsc{Poor description} (60\%): Students pointed out that the main challenge in replicating the authors' results was a poor description of the process. This issue includes missing details about the parameters used in the modeling (\eg, size of the random forest model), little information on the data preprocessing steps, inconsistency between the data released and the description in the article, and explicit mistakes of the authors in reporting the method details (\eg, wrong start date in a time series analysis). This issue was summarized by one student as: \say{It's almost a guessing game as to what method or inclusion I might be doing differently. This lack of hints was fairly difficult to navigate.}

\vspace{3mm}
\noindent \textsc{Expertise requirements} (37\%): 
Many students mentioned their lack of expertise as one challenge they encountered during the replication. Their descriptions varied from specific issues, such as the need to be confident in manipulating and plotting the data (\eg, how to plot timestamps on the \textit{x}-axis), to more complex problems, such as the use of some advanced techniques (\eg, domain-specific hypothesis testing).

\vspace{3mm}
\noindent \textsc{Time requirements} (17\%):
Students frequently mentioned the amount of time they spent working on the replication as a challenge. This problem is often associated with a poor description and is often described as many trial and error attempts.

\vspace{3mm}
\noindent \textsc{Limited resources} (11\%):
Finally, some students found working with the data provided challenging because of its scale. The computation time required to process large data sets represented a limitation for students working with personal laptops.

\xhdrNoPeriod{What may explain the differences between the original and the reproduced result?}
In this question, we investigate what the students believed could explain the differences between the figure in the paper and the one they obtained in the replication task. First, we asked as a multiple-choice question if they were able to replicate the results exactly (a), qualitatively (b), or not at all (c). Then, students were asked to describe in two or three sentences what may explain the differences.
The most common outcome is that figure ``replicated qualitatively but not exactly'' (b, 73.2\%), followed by ``did not replicate at all'' (c, 13.9\%), and ``replicated exactly'' (a, 12.7\%).

In this second exploratory analysis, we focus on students who obtained similar results (b) or failed to reproduce the figure assigned (c).
We identified five recurrent topics mentioned by the students who could not replicate the figure exactly: \textsc{Poor description}, \textsc{Data issues}, \textsc{Authors' mistakes}, \textsc{Tools differences}, and \textsc{Students' skills}. As for the previous analysis, each answer can mention multiple problems. We found that 83\% of the answers are assigned to at least one theme, while the remaining 17\% were not informative and could not be assigned to new topics.

\vspace{3mm}
\noindent \textsc{Poor description} (55\%):
Similarly to what we observed in the answers to the previous question, students blame the limited description for the mismatch between their results and the article's figure. Answers in this category frequently mention a lack of details on the models' parameters used by the authors. Students who managed to reproduce the results only qualitatively pointed out that it was impossible to reproduce the figure exactly when the code and seeds used for `random initialization' are unavailable. Another common observation was the limited description of all the steps and choices involved in the preprocessing pipeline. These aspects include how authors sampled data, handled missing values, what qualifies as outliers, and what numeric rounding steps are involved.

\vspace{3mm}
\noindent \textsc{Data issues} (30\%):
Many students attribute their impossibility of reproducing the results to problems associated with the data. These problems come from issues with the data release that does not entirely match the description in the paper or from an incomplete release of the data necessary to reproduce all the results. Students encountering this last limitation went as far as trying to collect their own data set with all the associated challenges---especially if depending on an outdated automated programming interface (API).

\vspace{3mm}
\noindent \textsc{Authors' mistakes} (24\%):
A significant portion of students assigned the blame for the impossibility of reproducing the results to the authors of the research. Answers assigned to this category mentioned possible embellishment of the results by the researchers and both genuine mistakes in reporting or plotting (\eg, \say{The authors interchanged a row at some point which messed up their analysis}) and bad-faith adjustments (\eg, \say{The authors did some shady-ish things, for example hard coding the plot}).

\vspace{3mm}
\noindent \textsc{Tool differences} (11\%):
Some students suspected that the discrepancy between the tool used for the replication may play a role in obtaining different results. They speculated on potential differences in the model and optimizer implementations available in Python, R, and Stata.

\vspace{3mm}
\noindent \textsc{Students' skill} (7\%):
Lastly, some students believe mistakes on their side can be a possible reason for the differences. Some of them mention general mistakes in their code, whereas others describe their inexperience in doing effective data preprocessing and using libraries or methods that are not explicitly covered in the course material (\eg, \say{Researchers used a very advanced algorithm from another paper and I would be surprised if any student fully implemented it.}). This topic is relatively infrequent likely due to the fact that, in preparation for the study, we identified publications suitable for the course, in terms of the difficulty of tasks required. We additionally ensured that the course lectures ahead of the replication covered the crucial skills necessary to perform the replication.

\xhdr{Paper-specific common feedback} Lastly, we aim to understand whether there were blocking factors that made it impossible for students to replicate the result that cannot be addressed simply by taking more time.\footnote{Note that our goal is \textit{not} to discover any specific replication failures, but rather to understand if the activity allows discriminating between potential true issues in the original publication from students' own issues.} We reexamined the students' explanations separately per paper in order to identify issues that students consistently mention when the result is not replicated. Such consistent issues that are reported many times might be authors' own mistakes or a true lack of information. 

We list recurring issues for five papers where more than 10\% of the students self-reported that they did not manage to replicate at all any of the assigned figures (\Tabref{tab:2}). Explanations for the remaining five papers did not contain any repeatedly occurring explanations. 

Examining the recurring explanations, we identified two recurring issues---a cross-validation mistake~\citep{muchlinski_comparing_2016} and counting error \citep{leskovec_signed_2010}, which were known to the instructors in advance and were correctly identified by students, while the other recurring issues mainly reflect a lack of information or other preprocessing discrepancies.

\begin{table}[t]
\centering
\scriptsize
\begin{tabular}{p{0.22\linewidth} | p{0.12\linewidth} | p{0.15\linewidth}| p{0.2\linewidth}| p{0.15\linewidth}}

                \multirow{2}{*}{ \textbf{Paper}} & \multirow{2}{*}{\textbf{Figure}} & \textbf{Replicated}&  \textbf{Replicated} & 
                 \textbf{Did not} \\
                                  &  & \textbf{exactly} &  \textbf{qualitatively} & 
                  \textbf{replicate}\\
                  \toprule
\multirow{2}{*}{ \cite{muchlinski_comparing_2016} } & Fig.~2 & 3.7\% & 96.3\% & 0\% \\
               &  Fig.~4   & 0\% & 55.56\% & 44.44\% \\
\multicolumn{5}{l}{Random forest parameters and random seed are not stated in the paper.}          \\
\multicolumn{5}{l}{Specific feature importance metric is not explained in detail.}          \\
\multicolumn{5}{l}{Programming language or library incompatibility meant that students could not reproduce the order }          \\
\multicolumn{5}{l}{of features. Authors' mistakes in the cross-validation procedure.}          \\
\midrule
\multirow{2}{*}{\cite{cho_friendship_2011}  } &  Fig.~2A & 7.69\% & 80.77\% & 11.54\% \\
                  & Fig.~3B & 0\% & 84.62\%  & 15.38\% \\
\multicolumn{5}{l}{Outlier removal and the prepossessing steps are not explained in sufficient detail.}          \\
\multicolumn{5}{l}{Null baseline is not explained in detail.}\\
\midrule
\multirow{2}{*}{ \cite{leskovec_signed_2010} } & Tab.~1 & 6.9\% & 93.1\% & 0\% \\
                  & Tab. 3 & 6.9\% & 55.17\% &  37.93\%\\
\multicolumn{5}{l}{Authors' mistakes in data processing and counting.}          \\
\midrule
\multirow{2}{*}{\cite{aiello_tesco_2020}} & Fig.~5 & 81.82\% & 18.18\% & 0\% \\
                  & Fig.~4 & 4.55\% & 59.09\% & 36.36\% \\
\multicolumn{5}{l}{Not sufficient information provided in order to reproduce the figure.}          \\
\multicolumn{5}{l}{Scaling of the distributions not explained in detail.}          \\
\midrule
\multirow{2}{*}{\cite{penney_chilling_2016} } & Fig.~3 & 0\% &  96.43\%& 3.57\% \\
                  & Fig.~4A & 0\% &  89.29\% & 10.71\% \\
\multicolumn{5}{l}{The original data set is not available. The data set that the students used contained slight discrepancies.}         
\end{tabular}
\caption{\textbf{Challenges students encountered, separately by paper (exploratory analysis).} The frequency of self-reported outcomes across students. For each paper, specific shared challenges that the students identified in their open-form responses when asked to explain the discrepancies between the original result and their result, and to speculate as to why the differences arose. Note that qualitative replication is the most frequent self-reported outcome for each of the figures.}
\label{tab:2}
\end{table}

\section{Results: Creative replication extensions}
\label{creative}

{As part of Step 5, in their groups, students conducted creative extensions of the analysis performed in the paper. According to the instructors' anecdotal experience, this creative component of the project---which students built on top of the replicated papers---was in many cases more technically advanced and meaningful compared to the unconstrained project in regular iterations.}

{To confirm these observations, we conducted several follow-up exploratory analyses. We analyzed structured project descriptions provided by the students in a consistent format. Across class iterations, these were submitted at the start, and updated at the end of the project. The descriptions were provided in a structured \texttt{README.md} document and contained a title, abstract, and a description of the research question(s), data set(s), and methods.}

{First, we developed a systematic method to automatically code structured project descriptions for the type of approaches each one uses and their scientific contributions (using GPT-4 model; Appendices, Appendix C: Annotation Details), an approach evaluated for similar text classification tasks~\citep{doi:10.1073/pnas.2305016120}. Specifically, we developed two custom preprompts for GPT-4 and applied each preprompt to the structured project descriptions that students are required to write. We confirmed that the GPT-4-generated annotations had high agreement with independent human annotations on a subset of descriptions. In particular, two authors annotated a random sample of project descriptions. The sample was also annotated leveraging the GPT-4 model, using the same instructions (see Appendices, Appendix C: Annotation Details for specific prompts and model parameters). We measured a substantial agreement between the manual and automated annotations ($\kappa=.70$ and $\kappa=.77$). Complete instructions and details about the agreement metrics are listed in Appendices, Appendix C: Annotation.}

{Then, we applied this method to the structured project descriptions. Each project description was annotated to indicate what type of data analysis method the project leveraged, among those covered in the class. The methods ranged from simple descriptive approaches, over less simple approaches (inference and prediction), to more technically advanced causal inference techniques. Following the same approach, each project description was also annotated to indicate whether it is scientifically relevant, by considering whether the project potentially pushes the boundaries of current scientific knowledge, as adapted from National Science Foundation definition of transformative research~\citep{nsf_2024}.

{We annotated descriptions of replication extensions conducted after data analysis replications performed as part of the project component of the class (2020, $N_1=115$), and descriptions of standard, open-ended projects, which were conducted in the following year (2021, $N_2=114$), when data analysis replications were not integrated into the class, but students instead independently proposed and conducted a project topic of choice. Students had the freedom to select their own project topic such that it did not rely nor build on data analysis replication.\footnote{In the year preceding the data analysis replication exercise (2019) the project description had a slightly different structure. We, therefore, opted to analyze 2020 and 2021, two iterations when the structure was consistent.} We then compared the results between the two years, contrasting replication extension projects with open-ended projects as conducted in other iterations of the class. }

\xhdr{Creative replication extensions are more technically advanced than unconstrained projects} 
{In line with our anecdotal experience, we found that creative replication extensions are significantly less likely to be focused on descriptive statistics and data visualization (e.g., simple statistical tests and correlations) compared to unconstrained projects (5.22\% \vs\ 20.35\%; ${\chi}^2= 11.58$, $p = 6.67\times 10^{-4}$). Simultaneously, replication extensions are more likely to focus on causal inference and counterfactual techniques (e.g., effect estimation and matching) compared to unconstrained projects (9.57\% \vs\ 0.88\%; ${\chi}^2= 8.70$, $p = 3.18 \times 10^{-3}$). No significant differences were observed for statistical modeling and inference (38.26\% \vs\ 41.59\%; ${\chi}^2= 0.21, p = 0.64$) and machine learning and prediction (46.96\% \vs\ 37.17\%; ${\chi}^2= 2.41, p = 0.12$). Complete histogram across the two years in visualized in \Figref{fig:creative}. These findings were robust to the inclusion of specific methods as examples, and to an alternative format where all data analysis types that apply are selected, as opposed to one that applies the most. This confirmed that the more advanced data analysis type---causal inference---is the least frequent analysis type, but more frequent among replication extensions than unconstrained projects (\Tabref{table:alternative}).}

{In summary, this exploratory analysis is aligned with the insight that creative replication extensions tend to be more methodologically advanced compared to unconstrained projects. Replication extension projects are associated with an increased use of more advanced causal data analysis methods, and a decreased use of less advanced descriptive methods.}

\begin{figure*}
\centering
    \includegraphics[width = \textwidth]{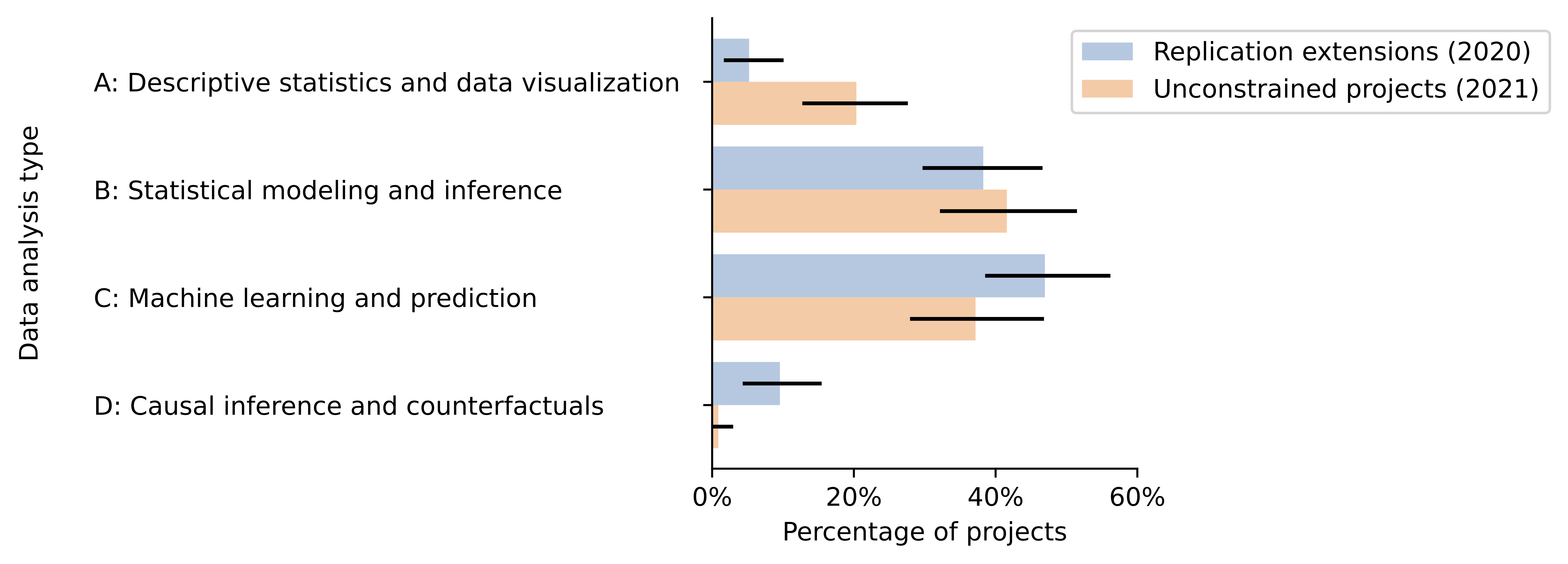}
\caption{{\textbf{Data analysis types, between years}. Histogram of the data analysis type across projects, in 2020, the year of creative replication extensions (blue), and 2021, the year on unconstrained projects (orange). Error-bars mark bootstrapped 95\% CI. Creative replication extensions are more technically advanced than unconstrained projects, as captured by a decreased use of less advanced descriptive methods (A), and an increased use of more advanced causal data analysis methods (D).}}
\label{fig:creative}
\end{figure*}

\xhdr{Creative replication extensions are more meaningful than unconstrained projects} {Compared to unconstrained projects, replication extensions were significantly more likely to be classified as scientifically relevant (15.65\% \vs\ 5.26\%; ${\chi}^2= 6.59$, $p = 1.03 \times 10^{-2}$), again confirming the insight that replication extensions are more meaningful than unconstrained projects.}

{Lastly, to explore whether projects differ in further ways, beyond those tested, we annotated project descriptions with adjectives that best capture the strengths of the project (Appendices, Appendix C: Annotation). We identified four adjectives occurring with a significantly different frequency between the two years ($p<.05$; full results in \Tabref{table:adjs}): ``practical'' and ``methodical'' (more frequent among replication extensions), and ``insightful'' and ``comprehensive" (more frequent among unconstrained projects). This analysis again points to unconstrained projects being more descriptive (comprehensive and providing insights), while replication extensions of replicated work focus on systematically executing advanced methods, and are of practical value and relevant (methodical and practical), as hypothesized.}

\xhdr{Summary} {Exploratory analyses point in the same direction as the instructors' experience---replication extensions are both more technically advanced and scientifically meaningful than unconstrained projects. Perhaps counterintuitively, creative extensions built on top of replications might be more meaningful than unconstrained projects~\citep{rosso2014creativity}. Unconstrained, students test many potential paths since they have not yet performed a viable data analysis. In contrast, extensions of data analysis replications allow going further beyond numerous shallow analyses, and are therefore more meaningful, allowing students to start from a strong foundation.}

{We note that this analysis is exploratory. Many other factors could contribute to differences between creative replication extensions and the more traditional, open-ended data analysis projects. These could include fundamental differences in the student body, instruction, and broader factors related to the class and the external environment.} 

{For instance, students performed projects using different data sets, and data set type could impact both the methodologies and the scientific relevance. However, in principle, different data sets allow performing all the data analysis types. When considering the data set type and limiting the projects only to those that primarily analyze the most common data set type (textual data, 113 in total), we still see consistent patterns such that the leveraging causal inference methods (5.88\% \vs\ 1.04\%) and scientific relevance (35.29\% \vs\ 5.21\%) are more common among replication extensions. Similarly, grading guidelines, instruction, and student prerequisites were otherwise unchanged. Nonetheless, other factors could impact these patterns and future work is needed to truly identify advantages of replication extensions compared to traditional assignments.}

\section{Considerations for educators}\label{sec:costs}

Having described how students experience data analysis replications, we now report insights and further `lessons learned' that can be useful to educators, with a particular emphasis on the necessary considerations to integrate data analysis replications into a class. The outlined points are based on the instructor and assistants' experiences and discussions, students' anonymous feedback, and the results of the surveys administered throughout the class. Although data analysis replications might have their advantages (\cf \Secref{creative}), integrating them into an existing course is challenging. Based on our study, we highlight five major considerations.

\subsection{Logistics} When designing and conducting in-class data analysis replications, it is necessary to carefully reevaluate and implement changes in the order in which the concepts are taught throughout the semester, since replicating data analyses requires specific skills (such as statistical tests, regression modeling, or counting items). One has to ensure that at the time when students start working on it, they have the required knowledge, which can lead to tradeoffs. In the study, in addition to modifying the class schedule, we carefully reconsidered other logistical aspects of the class, including group size and student assignment to projects and advisors.

\subsection{Human resources} In-class data analysis replication activities may require additional human resources. In our class (with $N=354$ consenting students), two teaching assistants dedicated half-time of their teaching assistantship to coordinating the project component of the course, as part of which the replication analysis was conducted. This amounted to around 8 hours per week. Additionally, around 30 students were assigned to each teaching assistant. The teaching assistants provided ongoing support specifically to the replicated paper throughout the semester, as well as performed grading, troubleshooting, technical support, and data analysis replications in preparation for the class.

\subsection{Added constraints} We note that, if implemented as part of a standard component of a class (\eg, project or homework), data analysis replications may constrain the topics, as students cannot perform a project of choice, but have to build on top of the data analysis replication. Additionally, student level needs to be considered, and the activity designed to be appropriate.
    
\subsection{Ethical challenges} Data analysis replication activities call for ethical consideration. First, we had doubts about assigning students to papers that we knew were likely not to replicate at all, because we did not want to give students tasks we knew were unlikely to succeed. The ethical issue of potentially knowingly exposing students to stress and frustration limits the pool of paper candidates. Second, since the replication activity takes students more time than expected, instructors should carefully plan the course timeline and communicate the expected workload clearly to students, to avoid any stress and frustration.

\subsection{Grading} {Grading guidelines were adapted to the replication exercise. Each student submitted a computational notebook containing well-commented code to create the figure or the table that was replicated, textual descriptions guiding through the process, and the figure/table that is the result of the replication.}

{Grading was independent of the replication outcome. Students were instructed that they would be graded based on the overall quality of the replication, textual descriptions, and code. It was noted that it would not be graded whether or not students managed to replicate the results from the paper, but only whether they had made an honest and diligent attempt at replicating, given the information available in the paper. We developed grading guidelines that specified the mapping between grades and the quality of textual descriptions and code, and provided graders with examples, which helped reduce subjectivity.}

Moreover, we advise caution in grading when assigning multiple papers within the same class. The selection of papers such that they are of comparable difficulty with regard to reproducibility is challenging, given that there are many paths one could take during a data analysis, and students are bound to face challenges that were not anticipated \citep{merrill_multiverse_2021}. Students are sensitive to a perceived uneven workload across groups, might prioritize performance in class, and in other ways feel that it is unfair that there is variance across groups in the amount of time they had to spend. In our study, this was the only aspect of the data analysis replication activity that the students reflected on negatively in the anonymous feedback. Alternatively, a single classwide project would address the issue of an uneven workload, but might not fit specific students' interests.


\section{Discussion}
\label{sec:discussion}

Our study characterizes students' experiences performing data analysis replications and derives insights and necessary considerations for educators aiming to incorporate them into classes.

\xhdr{Data analysis replications from the students' perspective} {First, testing our primary preregistered hypothesis, we found a significant difference between the expectations and reality of data analysis replications (\Figref{fig1}). The activity was more time-consuming and less challenging than anticipated, likely because the tasks were laborious and iterative (\Secref{exploratory}). It is noteworthy that the attitude shifts extended beyond the specific papers that the students replicated, into control papers, where following the initial replications, students' expected time to perform the replication increased by about two hours (\Figref{tab:h4}). The identified discrepancies between expectations and reality, and the observed changes in expectations about the reproducibility, serve as evidence of students' attitude shifts that have the potential to promote students' appreciation for the challenges involved in the scientific process.}

{Second, the creative component of the project, which students built on top of the replicated papers, was more technically advanced and meaningful than what students do in a fully unconstrained project in regular iterations, according to the instructors' experience and exploratory analyses of produced artifacts. This implies that data analysis replications might serve as one way to prepare students for addressing methodologically advanced and scientifically relevant problems.}

\xhdr{Data analysis replications from the educators' perspective} {Integrating data analysis replications into an existing course requires thoughtfulness and can run into challenges. We outline essential considerations for educators. Overall, we emphasize the need for careful logistics planning, allocating sufficient human resources, addressing ethical challenges, and devising appropriate grading strategies. We advise grading based on effort and methodology rather than replication outcomes. Moreover, we highlight and discuss necessary adjustments in course design, including the sequence in which concepts are taught and group sizes. We strongly emphasize the need for appropriate teaching assistants to support students and manage workloads, alongside carefully considering and selecting in advance publications that match both students' skill levels and individual interests.}

\xhdr{Data analysis replications from the scientists' perspective} {Moving forward, the scientific communities could potentially benefit from this and similar efforts. Teaching students to do data analysis replications can increase the overall number of conducted replications. Further advantages include a potential shift of norms and incentives if the auditing paradigm becomes more prevalent. If researchers are aware of large data analysis replication attempts and more replications are done, more attention may be paid to reproducibility in the future. Lastly, students' own experiences with replications may have an impact on their understanding and appreciation of reproducibility problems, and lead them to take measures to ensure that their own work is reproducible.}

\subsection{Limitations}

{We note that we are not measuring how replication exercises prepare students to practice computational science. Are replication exercises effective in teaching coding skills, deepening understanding, or gaining confidence in conducting independent research? While our study does not address these questions, we paint an initial picture of how students experience data analysis replications, and how that experience enhances students' understanding of what a data analysis replication entails.}

Similarly, our study does not disentangle the educational impact of a data analysis replication task from the educational impact of another comparable data analysis task. We contrast measurements before and after the activity, without randomly assigning students to the experimental conditions. Randomized assignment to the replication activity \vs\ another type of data analysis activity was considered but ruled out due to ethical challenges and to avoid student frustration. On the contrary, self-selection into a condition (replication \vs\ standard data analysis) would introduce biases and was hence also ruled out. Nonetheless, carefully designed cross-sectional longitudinal comparisons (\Secref{sec:methods}) can help tease out the impact of within a set of students who are all performing the replication activity, without contrasting to other data analysis tasks.

\subsection{Future work}

Our study opens the door for a number of future directions aiming to understand how to conduct in-class replication activities. First, exploratory analyses of students' perceptions of the ability to reproduce revealed a tension between attributing inconsistencies either to the authors' mistakes or the students' mistakes due to a perceived lack of skill. This raises an interesting follow-up question of determining the moment when a replication attempt can be considered complete and a student can stop performing the exercise, as opposed to assuming the inconsistencies can be attributed to the students' (lack of) skills or mistakes. It remains unclear---what is considered a sufficient and satisfactory time investment? How to avoid having students commit an unlimited amount of time to unproductive replication attempts?

One proposed solution may involve providing students with a limited number of submissions to a platform for corrected checks. This can involve allocating a `budget' with the number of attempts submitted to a platform to evaluate the data analysis results (similar to leaderboards where participants submit the predictions on a test set for evaluation to a platform). This approach would, however, require the instructors to a priori know the correct results of the data analysis to be performed, which would in turn defeat the power of data analysis replications to serve as a detector of false results in published papers.

Second, future efforts should consider building a crowdsourced cohort of university students to standardize and unify similar efforts \citep{berkeley_2024,schooler_metascience_2014,magazine2017replicationwiki}. Such efforts to redesign undergraduate courses for reproducibility and collaboration across institutes can result in fostering open science \citep{button_reboot_2018}.

Third, our study was based on 10 preselected publications tested in advance. In the future, we envision development of an auditing paradigm where classrooms are fundamentally integrated into the scientific process to evaluate comprehensive samples of published scientific findings, beyond the carefully selected pool used here.

Finally, future research integrating tools to support replication attempts is called for, including the usage of software containers, cloud computing, and checkpoints. These tools make it possible to standardize the computing environment around each submission \citep{liu_successes_2019,hofman_expanding_2021}. Standardizing the computing environment becomes particularly relevant in the age of closed-access large language models increasingly used as part of data analysis and modeling pipelines.

\subsection{Conclusion}

Our study explores the paradigm of in-class data analysis replications with a double purpose: to teach students while testing science.
We show that incorporating replications tasks into the project component of a large data science class has the potential to 
establish and increase the reproducibility of scientific work
as a natural by-product of data science instruction. We hope this article will inspire further instructors to consider including data analysis replications in the syllabus of their classes.

\subsection{Ethics} This study was approved by the EPFL Human Research Ethics Committee. We obtained consent for using the produced materials and survey responses for conducting research. Students were able to opt out of their data being analyzed. Students were provided with an information sheet (Methods, ``Information sheet for students''). The analyzed data is anonymized. Furthermore, the students were informed that any survey analyses would be conducted only after the class had already finished and the grades had been formed.

\subsection*{Disclosure Statement}
The authors have no conflicts of interest to declare.

\subsection*{Data and Code Availability}
Anonymized responses and the analysis code necessary to reproduce the results are deposited, and publicly available~\citep{DVN/A6VMD9_2024}.


\printbibliography

\setcounter{section}{1}
\setcounter{subsection}{0}

\newpage
\section*{Appendices}

\setcounter{figure}{0}
\setcounter{table}{0}
\renewcommand{\thefigure}{A\arabic{figure}}
\renewcommand{\thetable}{A\arabic{table}}

\subsection*{Appendix A: Primary Hypotheses--statistics and data distribution visualization.} 

Below, we list details about statistical analysis of our collected variables. All statistical tests were run with preregistered significance level $p = 0.05$. The unit of analysis is a student.

\begin{figure*}[b]
\centering
    \begin{minipage}[b]{0.39\textwidth}
    \centering
    \includegraphics[width=\textwidth]{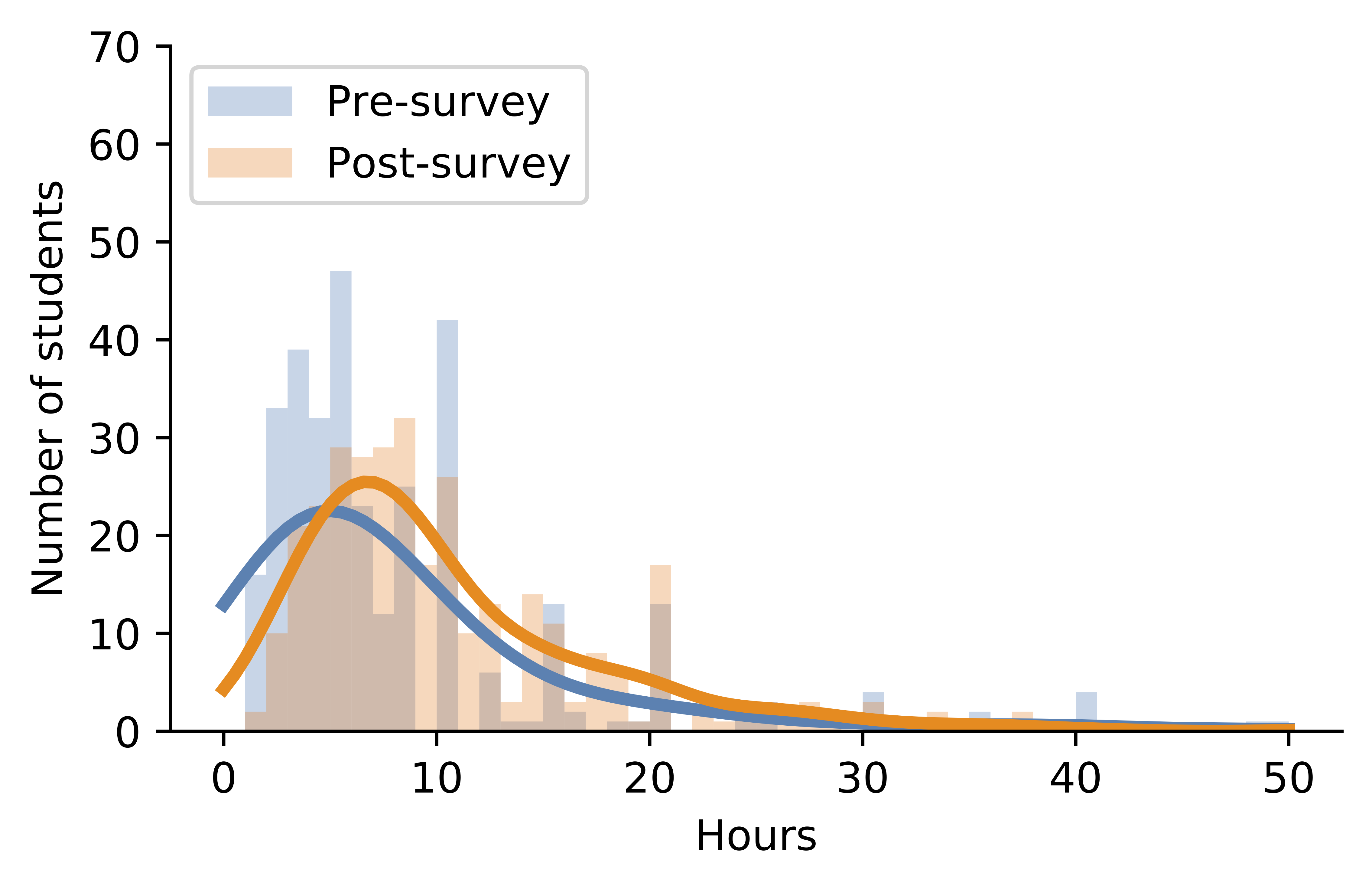}
    \subcaption{}
    \label{fig:H1a_1}
    \end{minipage}
  \hspace{1cm}
        \begin{minipage}[b]{0.39\textwidth}
    \centering
    \includegraphics[width=\textwidth]{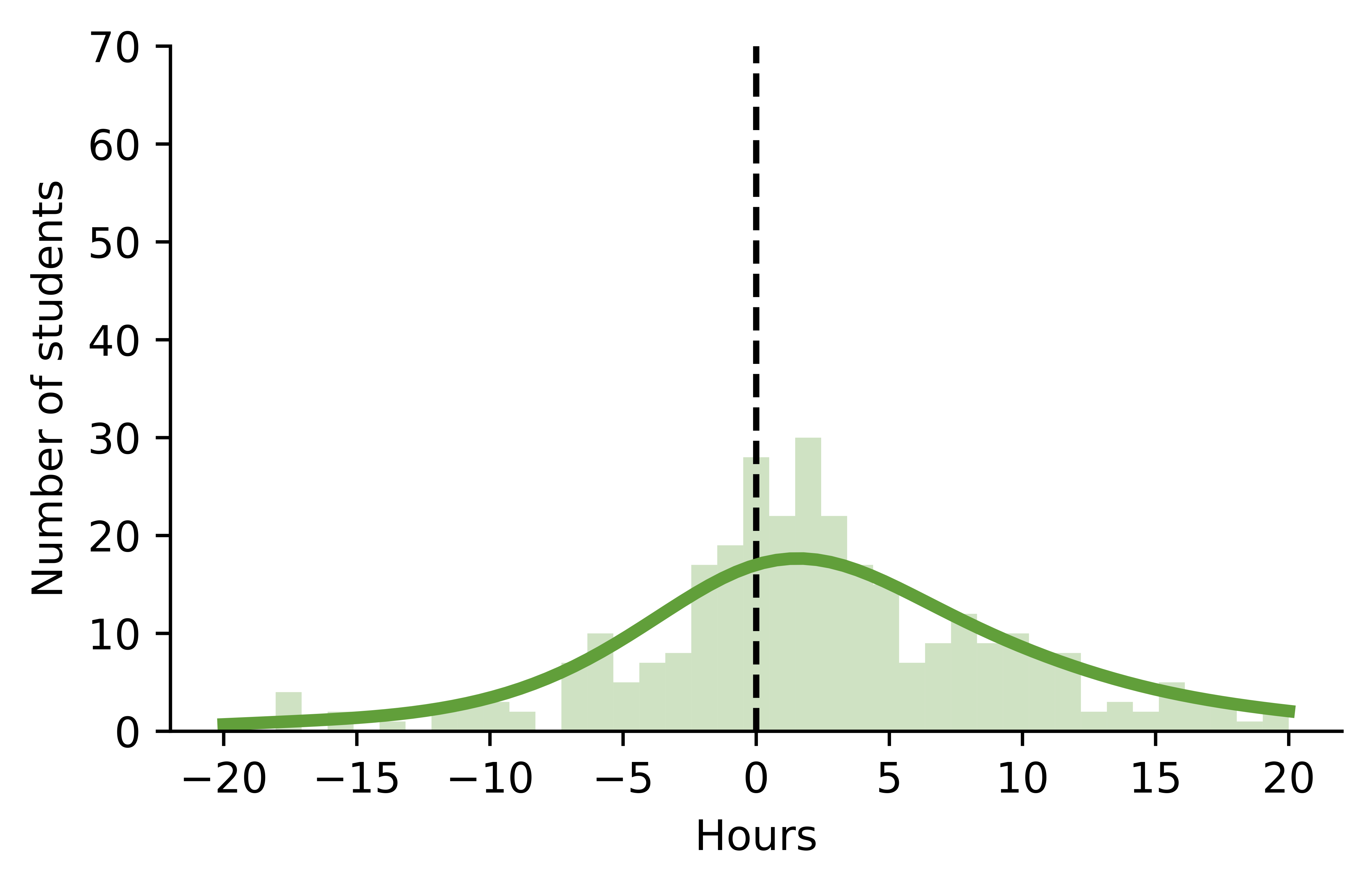}
    \subcaption{}
    \label{fig:H1a_2}
    \end{minipage}

\hspace{-0.3cm}
        \begin{minipage}[b]{0.39\textwidth}
    \centering
    \includegraphics[width=\textwidth]{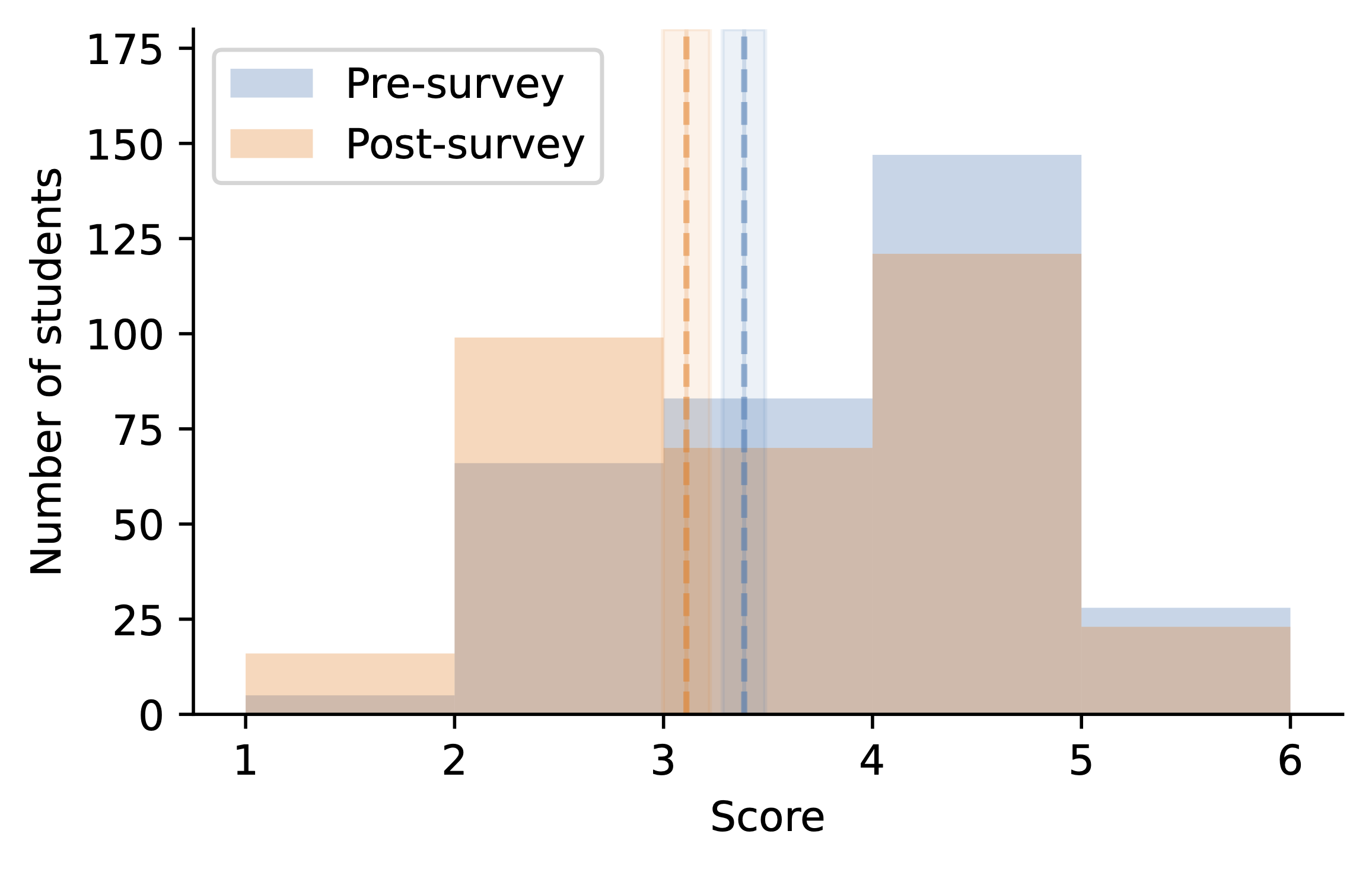}
    \subcaption{}
    \label{fig:H1b_1}
    \end{minipage}
  \hspace{1cm}
        \begin{minipage}[b]{0.39\textwidth}
    \centering
    \includegraphics[width=\textwidth]{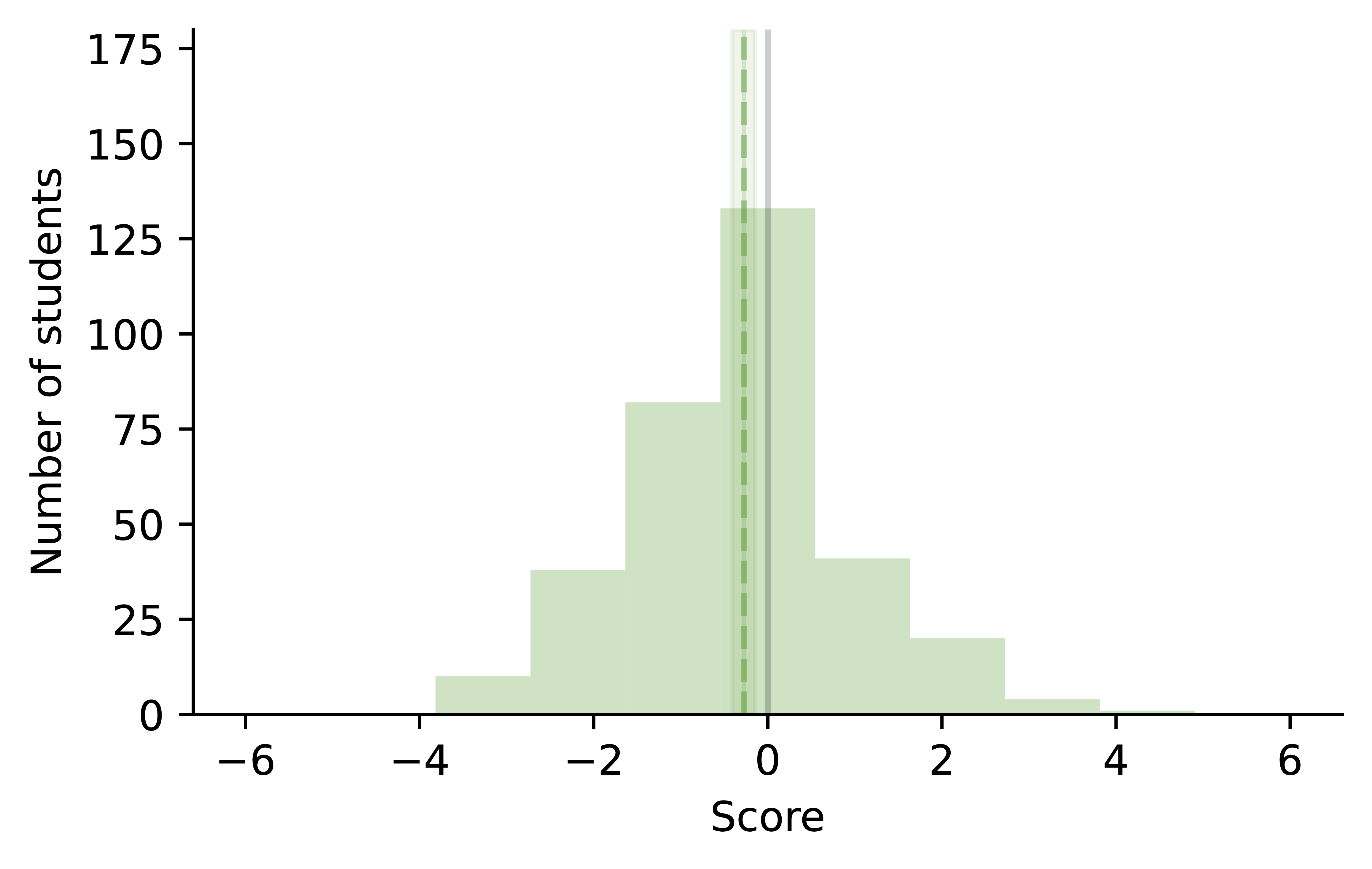}
    \subcaption{}
    \label{fig:H1b_2}
    \end{minipage}

    \begin{minipage}[b]{0.39\textwidth}
    \centering
    \includegraphics[height=3.8cm]{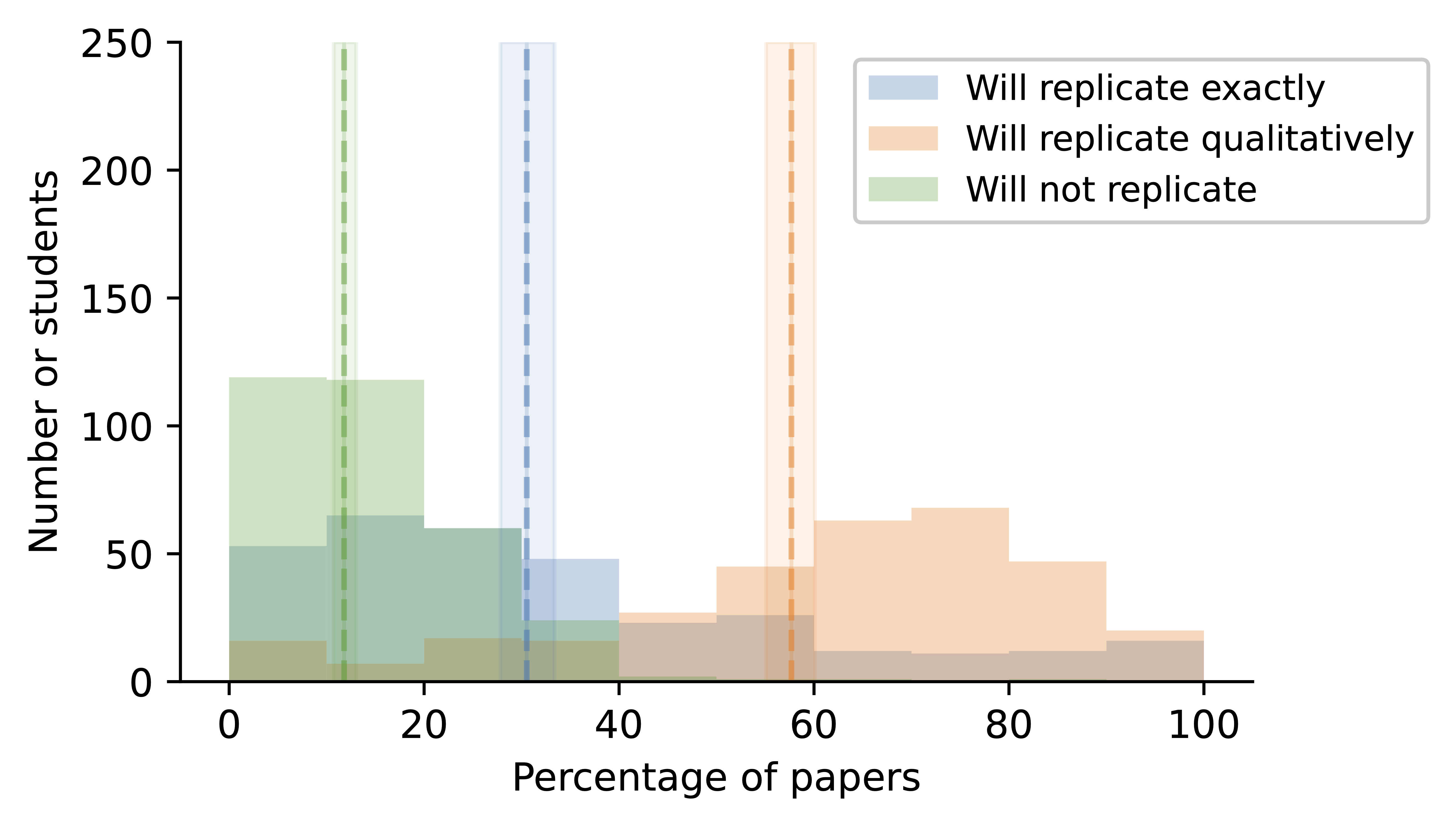}
    \subcaption{}
    \label{fig:H1d_1}
    \end{minipage}
  \hspace{0.8cm}
        \begin{minipage}[b]{0.39\textwidth}
    \centering
    \includegraphics[height=3.8cm]{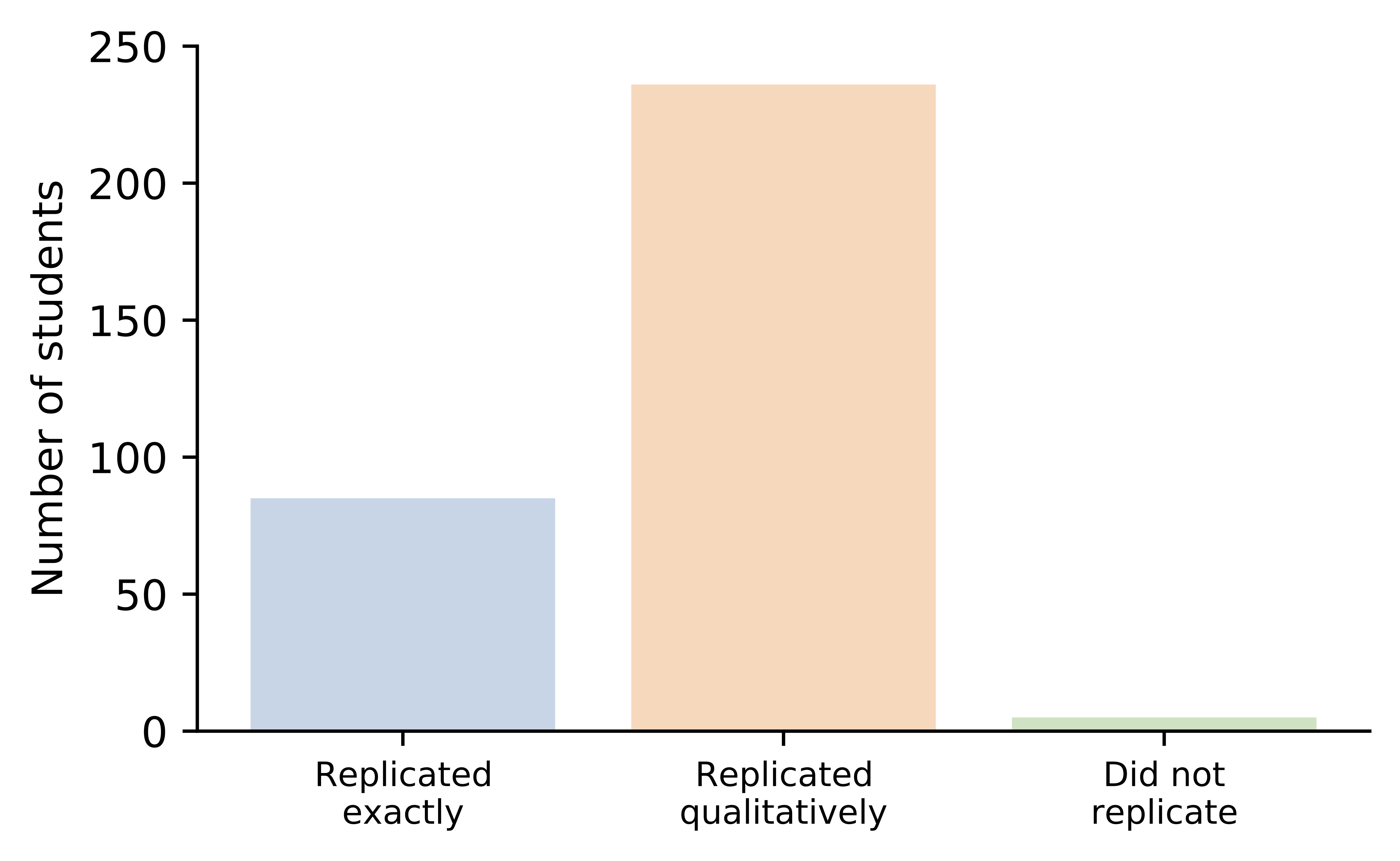}
    \subcaption{}
    \label{fig:H1d_2}
    \end{minipage}
    
\caption{\textbf{Expectations \vs\ reality of a data analysis replication exercise.} \textbf{Time taken (H1a).} (a) Across students (\textit{y}-axis), the histogram of the a priori expected number of hours (\textit{x}-axis) required (in blue), and the actual number of hours (in orange). (b) Across students (\textit{y}-axis), the histogram of the difference (\textit{x}-axis) between the actual number of hours and the expected number of hours. \textbf{Level of challenge (H1b)}. (c) Histogram of the expected level of challenge (on an ordinal 1--5 scale) of the data analysis replication in the presurvey (in blue), and the actual level of challenge of the data analysis replication (in orange). (d) Histogram of the difference between the expected and the actual level of challenge. The dashed lines and surrounding bands in each figure show the corresponding means and 95\% confidence intervals. {\textbf{Predicted and true outcomes of the replication (H1d)}. (e) Histogram of the percentage of papers expected to replicate exactly (blue), qualitatively (orange), or not at all (green), in the presurvey. The dashed lines and surrounding bands in each figure show the corresponding means and 95\% confidence intervals. (f) Histogram of true outcomes of the data analysis replication, in the postsurvey. 
}}
\label{fig1}
\end{figure*}

\noindent \underline{Preregistered analysis plan: Time spent} Students reported the expected number of hours in the presurvey and the actual number of hours in the postsurvey. Across students, we compared the expected number of hours to reproduce the basic figure from the assigned paper with the actual number of hours it took to reproduce. Specifically, we conducted a paired, two-sided $t$ test on the difference between actual and anticipated number of hours, with a null hypothesis of no mean difference.  

\noindent \underline{Preregistered analysis plan: Level of challenge} Students reported the perceived level of challenge on an ordinal scale (1: very straightforward, 2: somewhat straightforward, 3: neither straightforward nor challenging, 4: somewhat challenging, 5: very challenging). Specifically, we conducted a paired, two-sided $t$ test on the difference between the actual and anticipated level of challenge, with a null hypothesis of no mean difference.

\noindent \underline{Preregistered analysis plan: Distribution of time across core activities} Students were asked to sort three core activities with respect to the amount of time they expected to spend on them (before the analysis), and with respect to the amount of time they actually spent on them (after the analysis). The three activities can be ranked in six possible ways. We treated each of the six ranking configurations as a categorical variable. Our main hypothesis here relates to a disturbance in the rank of the three core activities. The ranking configurations in the pretest and the posttest were paired across students in a $6 \times 6$ contingency table. We then performed the Stuart--Maxwell test for marginal homogeneity in the contingency table. The null hypothesis is that the activity rank configuration frequencies for pretest and posttest are the same. 

\noindent \underline{Preregistered analysis plan: Replication outcomes} There are three possible self-reported outcomes of the data analysis replication: {the analysis replicated exactly} (the replication attempt produced results that agreed exactly with the paper, up to the decimals printed in the paper or shown in the figures), {the analysis replicated qualitatively} (the replication attempt produced results that had small differences with the paper, but these still agreed with the abstract-level findings of the paper), and {the analysis did not replicate at all} (the replication attempt produced results that were in conflict with the abstract-level findings of the paper).

We considered these outcomes as ordinal variables (1: the analysis replicated exactly, 2: the analysis replicated qualitatively, 3: the analysis did not replicate at all). In the presurvey, students attributed a probability to each of the possible outcomes. We calculated the outcome expectation on the ordinal scale for each student, by multiplying each possible outcome (1, 2, and 3) with the probability the student attributed to it and summing up. In the postsurvey, students selected one of the outcomes. We compared the anticipated and the true value across students, for the basic figure from the assigned paper. We performed a paired two-sided $t$ test.

\newpage

\setcounter{figure}{0}
\setcounter{table}{0}
\renewcommand{\thefigure}{B\arabic{figure}}
\renewcommand{\thetable}{B\arabic{table}}

\subsection*{Appendix B: Secondary hypotheses.}

Here we provide detailed statistics and analyses addressing a set of secondary hypotheses ({H2--4}) summarized in the main text.

\noindent \underline{H2 (RQ2):} \emph{Discrepancies between predictions and true outcomes persist as students solve replication tasks.}

In the first replication task, the students replicated the basic figure, and in the second replication task, they replicated the advanced figure. We compared the predictions and true outcomes for the advanced figure in the assigned paper by repeating the same analyses and statistical tests described in H1a--d, but now for the advanced rather than the basic figure. We then explored the ways how the second replication task differs from the first replication task. In other words, we explored how the discrepancies between expectations and outcomes vary as students gain experience in conducting data analysis replication tasks.

\begin{enumerate}[label=\alph*)]
    \item We found that a significant difference between the time students take to perform the advanced data analysis replication and the time they expect to take ($p=.0311$). On average, students expected to take 9.43 hr and took 8.54 hr. That is, after underestimating the time it takes to reproduce the basic figure, students overestimated the time it would take to reproduce the advanced figure (\ie, students overshoot after they initially underestimated).

    \item We found that, after performing the replication of the basic figure, there was a significant difference between how challenging performing data analysis replication of the advanced figure is, and how challenging students expect it to be. Performing data analysis replication tasks was again less challenging than expected ($p=.00766$) as students overestimated how challenging it would be. The average expected score on a 1--5 scale is 3.10, whereas the average score after performing the task is 2.91. For comparison, in the case of the basic figure, the average expected score on the 1--5 scale was 3.39 and the average score after performing the task was 3.11. 

    \item For the advanced figure, we again found discrepancies between the predicted and the true distribution of time spent on the three core activities: data wrangling, data analysis, and interpretation ($p=9.54\times 10^{-7}$). In particular, on average, data wrangling and data analysis took less time than expected, while interpreting results took more time than expected. For the advanced figure, students again overestimated how much time data wrangling would take, and underestimated how much time interpreting the results would take. We find no significant difference for the data analysis component.

    \item For the advanced figure, we found discrepancies between predicted and true outcomes of the replication ($p = 1.17\times 10^{-13}$). As a reminder, we considered these outcomes as ordinal variables (1: the analysis replicated exactly, 2: the analysis replicated qualitatively, 3: the analysis did not replicate at all). The pretest average score is on average 1.76, whereas the posttest average score is 2.01. Overall, the outcomes were less successful than expected. That is, with the advanced figure, students faced more reproducibility issues than with the basic figure, as we expected.
    
\end{enumerate}

\noindent \underline{H3 (RQ3):} \emph{The replication task affects the students' expectations on the fraction of peer-reviewed data science papers that are reproducible.}

At the beginning and at the end of the study, we asked the following question:
``Out of 100 peer-reviewed data science papers published in 2020, in how many of these papers do you think the analysis would replicate exactly, the analysis would replicate qualitatively, and the analysis would not replicate at all?''

As before ({H1d}), we considered the outcomes as ordinal variables (1: the analysis replicated exactly, 2: the analysis replicated qualitatively, 3: the analysis does not replicate at all). We calculated the outcome expectation on the ordinal scale for each student, by multiplying each possible outcome (1, 2, and 3) with the probability the student attributed to it and summing up. Students were instructed to carefully verify that the three numbers add up to 100, and we excluded students whose responses do not pass this validation check. We then performed a paired two-sided $t$ test on the outcome expectation at the beginning and at the end of the study. We did not find evidence that the replication task affects the students' expectations of the fraction of peer-reviewed data science papers that are reproducible ($p=.143$; illustrated in \Figref{fig3}).

\begin{figure*}
\centering
        \begin{minipage}[b]{0.39\textwidth}
    \centering
    \includegraphics[height=4.5cm]{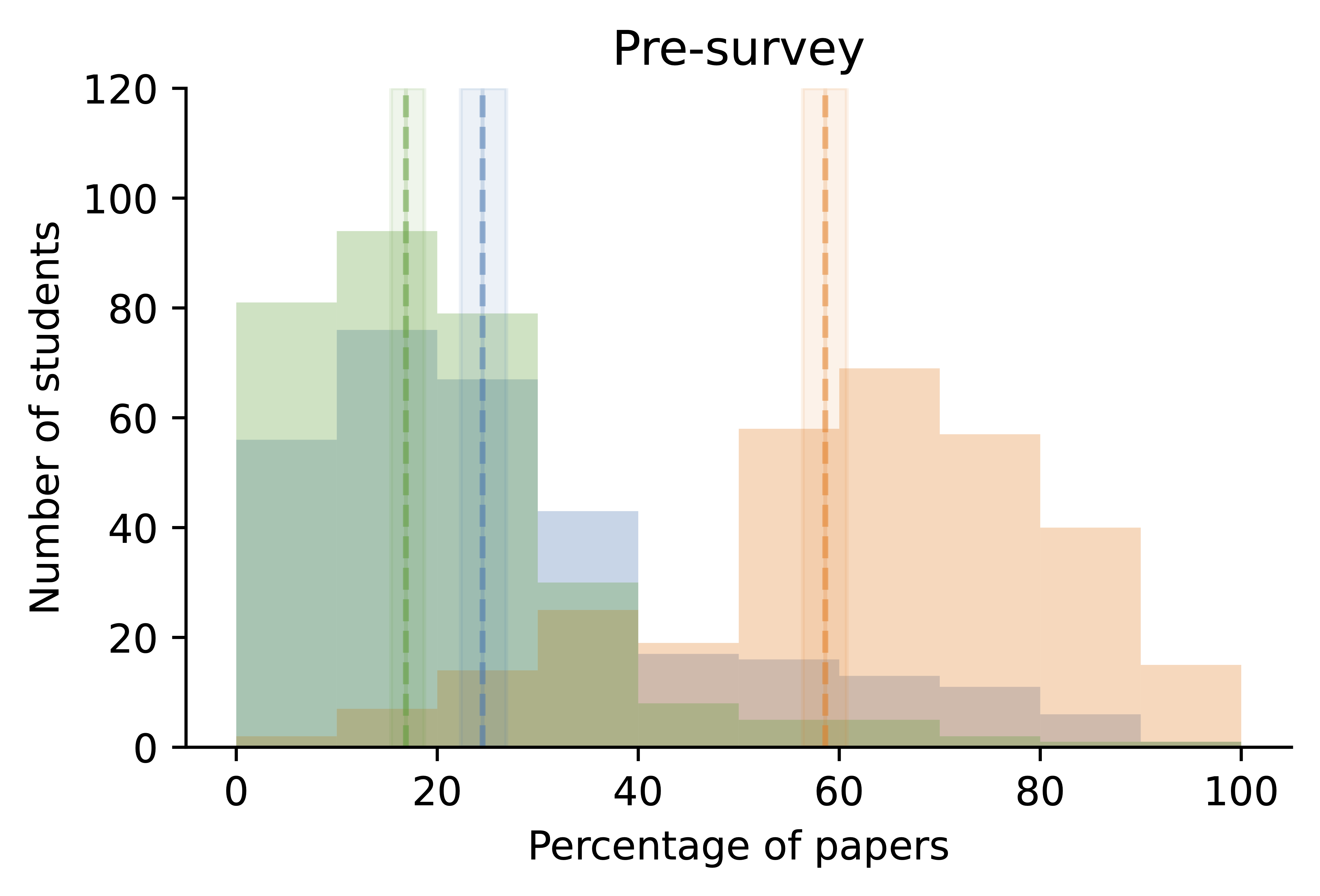}
    \subcaption{}
    \label{fig:H3_1}
    \end{minipage}
  \hspace{1cm}
        \begin{minipage}[b]{0.39\textwidth}
    \centering
    \includegraphics[height=4.5cm]{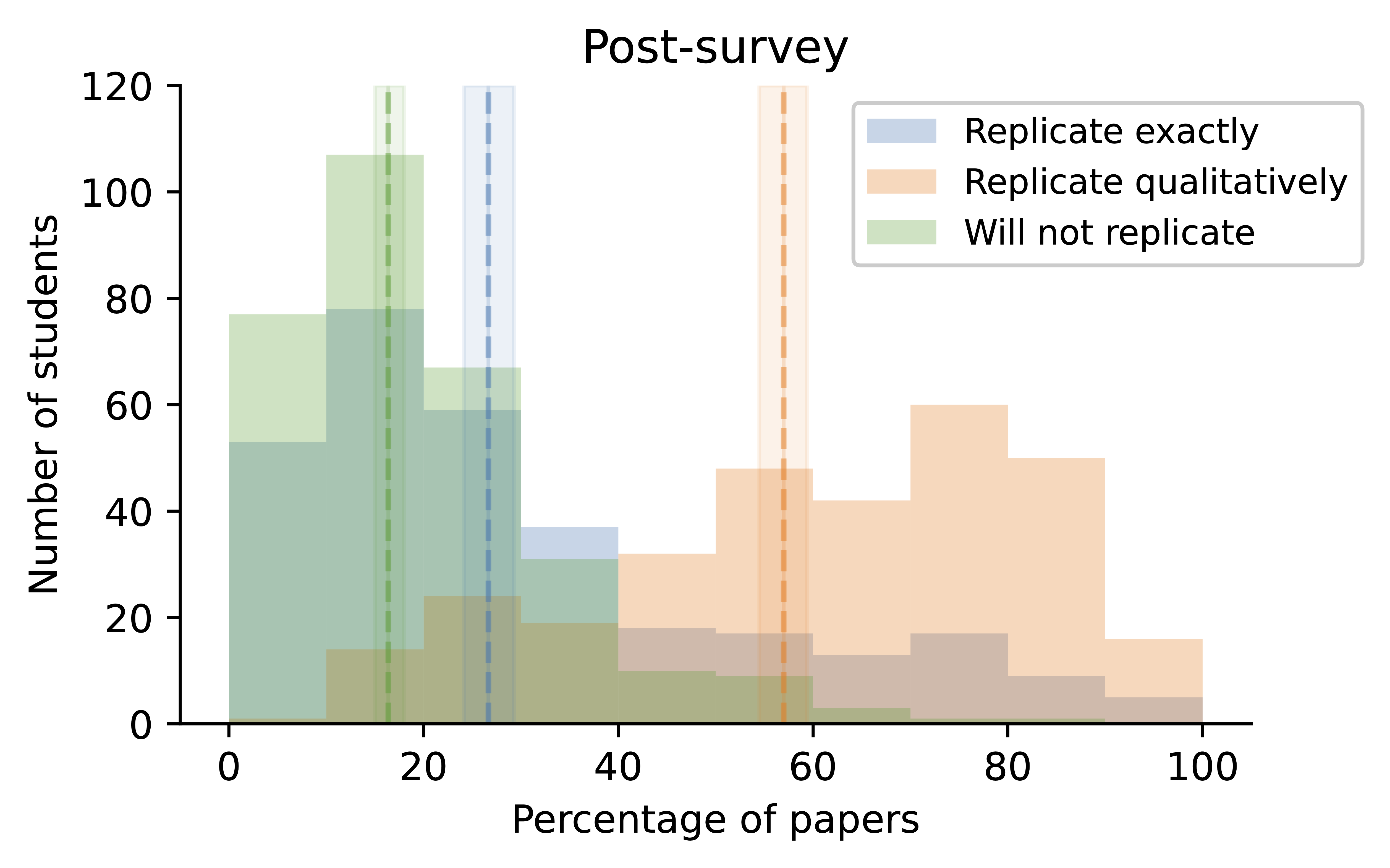}
    \subcaption{}
    \label{fig:H3_2}
    \end{minipage}
\caption{ \textbf{Perceived reproducibility of peer-reviewed data science papers (H3)}. Histogram of the percentage of papers expected to replicate exactly (blue), qualitatively (orange), or not at all (green), (a) in the presurvey, (b) in the postsurvey. The dashed lines and surrounding bands in each figure show the corresponding means and 95\% confidence intervals.}
\label{fig3}
\end{figure*}

\noindent \underline{H4 (RQ4):} \emph{There is a spillover effect as expectations are modified across the board, to papers that students did not replicate.}

Upon performing the replication tasks, we monitored any simultaneous changes in the expectations (as described in H1a--d) for the two control papers that students did not reproduce. We explored whether there were any changes in expectations regarding quantities described in H1a--d by repeating the same tests as outlined above, for the two control figures. One of the two control papers (`Paper 2') entails data analysis of the same type as the replicated paper (`Paper 1') and the other (`Paper 3') one of a different type (counting items and hypothesis testing \vs\ regression modeling). By contrasting the two control papers, we explored the presence of any spillover effects to different types of data analysis replication, beyond the specific type of analysis that the student worked on.

Overall, we found that there is a spillover effect as expectations regarding time spent and time distribution across the activities are modified across the board, for the papers that students did not replicate (summarized in \Tabref{tab:h4}). It is noteworthy that, even though the figures students were asked about were not replicated, the expectations changed after vs.\ before the replication activity. The expectations were modified in the same direction as for the replicated papers (by about 2 hours' increase in expectation, and more time expected to spend in analysis and interpretation). The effects are not stronger for the same type of data analysis as performed in the replication exercise. We found that, overall, there was an attitude shift across data analysis types.

\begin{table}[t]
\scriptsize
\centering
\renewcommand{\arraystretch}{1.5} 
\begin{tabular}{l|l}
\multicolumn{2}{l}{\textbf{Control paper of the same type as replicated}} \\
\midrule
H4a: Expected time                                & 1.73 hour increase ($p=.0129$). \\
& Pre test: $M=8.81$, post test: $M=10.54$.                                                                                    \\
\hline
H4b: Expected level of challenge                  &  Difference not significant ($p=.836$).                                                                              \\
\hline
H4c: Expected distribution  & Significant disturbance in the ranking ($p<10^{-307}$).\\
& Wrangling: +0.36, Analysis: -0.04, Interpretation: -0.32.  \\
\hline
H4d: Expected outcomes                            & Difference not significant ($p=.0804$).             \\                   \toprule
\multicolumn{2}{l}{\textbf{Control paper of a different type than replicated}} \\
\midrule
H4a: Expected time                                &  2.07 hour increase ($p=.000434$).\\
& Pre test: $M=8.75$, post test: $M=10.82$.                                         \\
\hline
H4b: Expected level of challenge                  &  Difference not significant ($p=.161$).                                                                              \\
\hline
H4c: Expected distribution  &  Significant disturbance in the ranking ($p<10^{-307}$). \\
& Wrangling: +0.37, Analysis: -0.07, Interpretation: -0.31. \\
\hline
H4d: Expected outcomes                             & Difference not significant ($p=.0841$).                                                                               
\end{tabular}
\caption{\textbf{Spillover effects: Modified expectations regarding papers that students did not replicate (H4).} Summary of the results comparing pre- and postreplication expectations, across the four hypotheses (H4a--H4c), for the two types of control papers that were not replicated.}
\label{tab:h4}
\end{table}

\newpage
\setcounter{figure}{0}
\setcounter{table}{0}
\renewcommand{\thefigure}{C\arabic{figure}}
\renewcommand{\thetable}{C\arabic{table}}
\clearpage
\newpage
\subsection*{Appendix C: Annotation details.}\label{section:annotation}

{Two authors independently annotated a set of 20 project descriptions (10 each). We calculated Cohen's kappa coefficient to probe interrater reliability between the authors and annotations conducted with GPT-4. For both questions (data analysis type and novelty of the scientific question), we found substantial interrater reliability between GPT-4 annotations and the authors' annotations ($\kappa=.70$ and $\kappa=.77$, respectively). Following this small-scale evaluation, we adopted automated annotation for exploratory analyses of project descriptions. We note that further evaluation is necessary to robustly validate this approach and extend it beyond the exploratory analyses described here.}

\noindent \underline{Prompting parameters.} {All the annotations were collected using OpenAI's ChatCompletion API endpoint. Model GPT-4 (`gpt-4-0613') was used, with default parameters (default temperature of 1). For reproducibility, we list the complete prompt texts below.}

\noindent \underline{Prompt text: How technically advanced is a project?} Which of the following types of data analysis applies to the described activities?\\
\noindent Select the one that applies the most.\\
\noindent A) Descriptive statistics and data visualization (e.g., Statistical tests, Correlation)\\
\noindent B) Statistical modeling and inference (e.g., Regression analysis, Logistic regression)\\
\noindent C) Machine learning and prediction (e.g., Predictive modeling, Clustering)\\
\noindent  D) Causal inference and counterfactuals (e.g., Effect estimation, Matching)\\
\noindent  Description: <README.md text>\\
\noindent Answer:

\noindent \underline{Prompt text: How scientifically meaningful is a project?} Is the proposed project pushing the boundaries of current scientific knowledge?\\
\noindent Answer YES or NO.\\
\noindent Description: <README.md text>\\
\noindent Answer:

\noindent \underline{Prompt text: Open-ended adjective generation.} List between one and five adjectives that best capture the strengths of this project. Focus on the questions, methods, results, and possible impact.\\
\noindent Output a comma-separated list of adjectives.
\noindent Description: <README.md text>\\
\noindent Answer:

\noindent \underline{Prompt text: Dataset type.} Which data type applies to the described activities?\\
Output a number corresponding to the most relevant data type.\\
1) Tabular data\\
2) Networks\\
3) Textual data\\
4) Other data types\\
\noindent Description: <README.md text>\\
\noindent Answer:

\begin{table}[ht]
\scriptsize
\begin{tabular}{lrllr}
\toprule
Does the data analysis type apply? &  Frequency, 2020 & Frequency, 2021 &${\chi}^2$  & \textit{p} value \\
\midrule
A) Descriptive statistics and data visualization & 6.96\% & 22.12\% & 10.41 & $1.26 \times 10^{-3} $\\

B) Statistical modeling and inference & 42.61\% & 46.90\% & 0.35 & .55\\ 

C) Machine learning and prediction &  42.61\% & 30.09\% & 4.05 &  $4.42 \times 10^{-2}$ \\

D) Causal inference and counterfactuals & 7.83\% & 0.88\% &  6.62 & $1.01\times 10^{-2}$ \\
\bottomrule
\end{tabular}
\caption{\textbf{Alternative annotation scheme}. Data analysis types, their frequencies across the two years, ${\chi}^2$ statistic, and the corresponding \textit{p} value. }
\label{table:alternative}
\end{table}

\noindent \underline{Adjectives.} Complete statistics for all the tested adjectives are listed in Table~\ref{table:adjs}.

\begin{table}[ht]
\begin{tabular}{lrrrr}
\toprule
Adjective &  Frequency, 2020 & Frequency, 2021 &${\chi}^2$  & \textit{p} value \\
\midrule
       \textbf{insightful} &     6.09\% &    12.81\% & 1513.19 & .000100 \\
       \textbf{practical} &     1.57\% &     0.18\% &  638.62 & .011501 \\
       \textbf{methodical} &     8.87\% &     5.44\% &  506.80 & .024372 \\
    \textbf{comprehensive} &    13.39\% &    18.07\% &  472.97 & .029647 \\
         detailed &     2.43\% &     3.86\% &  190.84 & .167144 \\
      informative &     0.87\% &     1.58\% &  119.27 & .274793 \\
      inquisitive &     0.17\% &     0.53\% &  102.11 & .312251 \\
   methodological &     0.52\% &     0.18\% &   98.61 & .320704 \\
     quantitative &     0.52\% &     0.18\% &   98.61 & .320704 \\
           robust &     0.52\% &     0.18\% &   98.61 & .320704 \\
    collaborative &     3.13\% &     2.28\% &   78.46 & .375728 \\
        impactful &     9.91\% &     11.40\% &   66.80 & .413761 \\
       innovative &    14.78\% &    13.16\% &   62.86 & .427885 \\
         relevant &     1.57\% &     2.11\% &   46.37 & .495906 \\
        strategic &     0.17\% &     0.35\% &   34.30 & .558086 \\
       systematic &     0.17\% &     0.35\% &   34.30 & .558086 \\
multidisciplinary &     0.35\% &     0.18\% &   32.55 & .568311 \\
      resourceful &     0.35\% &     0.18\% &   32.55 & .568311 \\
        inclusive &     0.35\% &     0.18\% &   32.55 & .568311 \\
         in-depth &     0.52\% &     0.35\% &   19.22 & .661088 \\
         rigorous &     1.04\% &     0.88\% &    8.32 & .773025 \\
         thorough &     1.91\% &     2.11\% &    5.37 & .816694 \\
       analytical &    14.09\% &    14.39\% &    2.10 & .884885 \\
      data-driven &     1.57\% &     1.58\% &    .03 & .985101 \\
 forward-thinking &     0.17\% &     0.18\% &    .00 & .995068 \\
  detail-oriented &     0.17\% &     0.18\% &    .00 & .995068 \\
           timely &     0.17\% &     0.18\% &    .00 & .995068 \\
       meaningful &     0.17\% &     0.18\% &    .00 & .995068 \\
\bottomrule
\end{tabular}
\caption{\textbf{Adjectives describing the projects}. Adjectives, their frequencies across the two years, ${\chi}^2$ statistic, and the corresponding \textit{p} value. Terms with statistically significant difference in frequency ($p<.05$) are marked in bold. For completeness, all the terms are listed.}
\label{table:adjs}
\end{table}

\end{document}